\documentclass[11pt,reqno,english]{amsart}
\usepackage{amsfonts,amsmath,latexsym,verbatim,amscd,mathrsfs,color,array,amssymb,amsthm,graphicx}
\usepackage{subcaption}
\usepackage[hmargin=2.0cm, vmargin=2.5cm]{geometry}
\usepackage{ stmaryrd}
\usepackage{xcolor}

\usepackage[overload]{empheq}   
\newtheorem{remark}{Remark}

\newtheorem{PR}{Main Result}

\newtheorem{lemma}{Lemma}

\newcommand{\bm}[1]{\boldsymbol{#1}}

\newcommand{\PD}[2]{\frac{\partial#1}{\partial#2}}

\newcommand{\paren}[1]{\left(#1\right)}

\newcommand{\mc}[1]{\mathcal{#1}}
\newcommand{\abs}[1]{\left\lvert #1 \right\rvert}
\newcommand{\ep}{\varepsilon}
\newcommand{\hot}{\mathrm{h.o.t.}}

\newcommand{\dist}{\mathrm{dist}}

\numberwithin{equation}{section}

\title[Sharp Interface NRCH]{Sharp Interface Dynamics in a Minimal Non-Reciprocal Cahn-Hilliard System}

\author[D. Gomez]{Daniel Gomez}
\address{\noindent Daniel Gomez, ~Department of Mathematics and Statistics, University of New Mexico, Albuquerque, NM USA 87131}
\email{danielgomez@unm.edu}

\author[Y. Mori]{Yoichiro Mori}
\address{\noindent Yoichiro Mori, ~Department of Mathematics, University of Pennsylvania, Philadelphia, PA USA 19104}
\email{y1mori@sas.upenn.edu}

\author[S. Strikwerda]{Sarah Strikwerda}
\address{\noindent Sarah Strikwerda, ~Department of Mathematics, University of Wisconsin-Madison, Madison, WI USA 53706}
\email{sstrikwerda@wisc.edu}

\begin{document}

\maketitle

\begin{abstract}
    Interest in non-reciprocally coupled systems recently led to the introduction of a minimal non-reciprocally coupled Cahn-Hilliard (CH) model by Brauns and Marchetti in 2024, which we refer to as the Brauns-Marchetti (BM) model. This model can be seen as a conservative counterpart to the spatially extended FitzHugh-Nagumo model. Lacking a gradient structure, the BM model was observed to exhibit interesting dynamics including traveling periodic wave-trains and other coherent structures, as well as spatiotemporal chaos in certain parameter regimes. In this paper, we derive an effective equation for the interface dynamics of solutions to the BM model in $\mathbb{R}^2$ in the sharp-interface limit. The resulting system of equations is a generalization of the classical Mullins-Sekerka (MS) equations, which we refer to as the modified MS equations. We show that the modified MS equation shares some properties with its classical counterpart, but importantly, it is not in general a length minimizing flow. To illustrate the utility of this asymptotic reduction in the sharp interface limit, we perform a detailed analysis of stationary and periodic wave-trains, systematically deriving expressions for wave-train speeds and stability thresholds. The methods used here should be applicable to other non-reciprocally coupled CH models and therefore provide another avenue for their more detailed analysis. 
\end{abstract}

\section{Introduction}
Dissipative systems that exhibit spatiotemporal pattern formation abound in both natural and artificial systems \cite{ frohoff_2021, fruchart_2021}. Given the sheer variety of such systems, it is useful to formulate canonical models that are simple enough to be analytically or computationally tractable yet retain some of the most salient pattern forming properties. One such model is the FitzHugh-Nagumo (FHN) model, which reads as follows
\begin{subequations}\label{FHN}
\begin{align}[left=\empheqlbrace]\label{FHN1}
\PD{u}{t}&=\ep \Delta u-\frac{1}{\ep}f(u)- \theta v,\\
\tau\PD{v}{t}&=-v-\tau^{-1}u,
\end{align}
\end{subequations}
where $\theta<0$ and $\tau>0$ are constants, and $f$ is the derivative of a double-well potential $f(u)=F'(u)$. This equation 
was originally formulated as a reduced model for action potential propagation along a neuronal axon \cite{FitzHugh1961, Nagumo1962}, and has since also found applications to many problems outside of neuroscience \cite{lacasa_20204}. The FHN model, in different parametric regimes, exhibits traveling waves, oscillations, spirals waves, and chaotic patterns. Its relative simplicity has led to an extensive study of the FHN model using both computational and analytic methods (see for example \cite{keener-2009} and the references therein).

Setting $\theta=0$ in \eqref{FHN}, the system decouples and \eqref{FHN1} becomes the Allen-Cahn (AC) equation
\begin{equation}\label{AC}
\PD{u}{t}=\ep \Delta u-\frac{1}{\ep}f(u).
\end{equation}
The AC equation can be seen as an $L^2$ gradient flow of the energy functional
\begin{equation}\label{energy}
\mc{E}[u]=\frac{1}{\ep}\int\paren{\frac{\ep^2}{2}\abs{\nabla u}^2+F(u)}dx.
\end{equation}
Moreover, the AC equation is a standard model for the study of propagating fronts, but cannot support oscillatory or chaotic patterns due to its gradient structure. The coupling with the second variable $v$  when $\theta< 0$ destroys this gradient structure which opens the possibility to richer dynamics.

Among pattern forming systems, conservative systems constitute an important subclass which can in general be written in the form 
\begin{equation*}
\PD{u_i}{t}+\nabla \cdot \bm{j}_i=0,
\end{equation*}
where $u_i$ is the species and $\bm{j}_i$ the flux for each $1\leq i\leq N$. The constitutive equations for $\bm{j}_i$ are given as functions of $u_1,\cdots, u_n$ and complete the system of equations.
If we regard $u_i$ as the concentration of a chemical species, the equations will generally have the above form if the species do not engage in chemical reactions. More generally, the dynamics of interacting agents that do not change type will generally be described by such conservative dynamics. Conservative systems exhibit dynamics that can be considerably different from that of non-conservative systems. With this in mind, it then becomes natural to seek a conservative counterpart to the FHN system. In \cite{brauns_2024}, the authors propose the following model
\begin{subequations}\label{BM}
\begin{align}[left=\empheqlbrace]
\label{BM1}
\PD{u}{t}&=-\Delta \paren{\ep \Delta u-\frac{1}{\ep}f(u)-\theta v}\\
\tau\PD{v}{t}&=\Delta(v+\tau^{-1}u)
\end{align}
\end{subequations}
where again $\tau>0$ and $\theta\neq 0$ are constants, and $f=F'(u)$ for a (well-balanced) double-well potential $F(u)$. Throughout the paper we will refer to \eqref{BM} as the {\em Brauns-Marchetti (BM) model}. The BM model can be formally obtained from the FHN model \eqref{FHN} by applying $-\Delta$, a self-adjoint positive semi-definite operator, to both equations. We also note that the BM model can be derived as a reduction of certain concrete physical models as discussed in \cite{brauns_2024}.

Setting $\theta=0$ in \eqref{BM} the two equations again decouple and this time \eqref{BM1} becomes the following Cahn-Hilliard (CH) equation
\begin{equation}\label{CH}
\PD{u}{t}=-\Delta \paren{\ep \Delta u-\frac{1}{\ep}f(u)},
\end{equation}
which can be seen as the $H^{-1}$ gradient flow of the energy functional \eqref{energy}. 
The CH equation is a canonical model for phase separation and coarsening, 
but it is limited in its capacity to generate more complex dynamic  patterns due to its gradient structure. The second equation in $v$ destroys this gradient structure, 
thus making it possible for the BM system to exhibit a greater variety of conservative spatio-temporal dynamics. Given the wide variety of dynamics shown by this equation, \cite{frohoff23} has argued that it should be considered an amplitude equation, expanding those described in \cite{Cross93}.

Indeed, in \cite{brauns_2024}, the authors report interesting patterns generated by the BM model \eqref{BM} including traveling pulses, undulating waves and chaotic patterns.
Their analysis however, is largely restricted to studying the perturbative behavior around a spatially homogeneous steady state and the formal extension of such methods to spatially inhomogeneous solutions. The goal of our paper is to present a mathematical analysis that systematically sheds light on the dynamics of the BM model beyond its behavior around spatial homogeneity.

We study the BM model in the sharp interfacial limit. Much of the understanding of both the AC and CH models are in this limit, which is obtained by letting the parameter $\ep\rightarrow 0$ in \eqref{AC} or \eqref{CH} \cite{rubinstein-1989,pego_1989}. In this limit, the solution to \eqref{AC}
and \eqref{CH} can be reduced to an evolution equation of the interface separating the two regions where the solution $u$
takes the two different values at the local minima of the double well potential $F(u)$. In the case of the CH equation \eqref{CH}, the resulting interfacial dynamics is governed by the Mullins-Sekerka (MS) model \cite{pego_1989, alikakos_1994}.
One of the main contributions of our paper is the derivation of a modified MS model for the BM model \eqref{BM} in the sharp interface limit $\ep\ll 1$. This reduction allows us to analytically construct spatially non-homogeneous stationary solutions and traveling pulse solutions. We can further study the stability of such solutions, thereby providing analytical insight into the emergence of oscillations
and undulating waves. We note that our development is analogous to how the dynamics of the FHN model has been clarified
by taking advantage of the sharp interface limit of the AC equation. 

Before proceeding to a description of our main results, we remark that the BM model is arguably the simplest among a class of models known as non-reciprocal CH models \cite{rana-2024,saha_2020, you_2020}. Recalling that both the AC and CH models are gradient flows of the energy functional \eqref{energy}, the thermodynamic interpretation 
is that they describe the dynamics of a system under relaxation whose free energy is given by \eqref{energy}. 
That this process can be described by a gradient flow is linked to the fact that relaxation dynamics must satisfy 
Onsager reciprocity \cite{doi_2011,wang_2021}. For non-equilibrium systems in which the system may receive internal or external energy input, the free energy will no longer decrease in time, thereby potentially violating reciprocity. A collection of recent papers has systematically explored different non-reciprocal extensions of the CH model. The techniques developed in this paper are expected to be applicable to this wider class of models.

\subsection{Main Results}\label{subsec:nrms}
In this paper, we derive a modification of the Mullins-Sekerka equation for the interface motion of the non-reciprocal Cahn-Hilliard equation proposed by Brauns and Marchetti \cite{brauns_2024} in the sharp interface limit. In order to more directly connect our results to the proposed model of Brauns and Marchetti, we first state their original system in dimensional variables:
\begin{subequations}\label{eq:brauns-marchetti-nrch}
\begin{align}[left=\empheqlbrace]
    & \partial_T\Phi = \Delta_{X}\left(-\kappa\Delta\Phi + \beta f(\Phi) + D_{12}\Psi\right), & X\in\widetilde{\Omega},\, T>0\\
    & \partial_T\Psi = \Delta_{X}\left(D_{22}\Psi + D_{21}\Phi\right), & X\in\widetilde{\Omega},\, T>0.
\end{align}
\end{subequations}
where $\widetilde{\Omega} = [0,L]\times[0,H]$. Introducing the non-dimensional variables
\begin{equation}
    T = \frac{L^3}{\sqrt{\kappa\beta}} t,\qquad X = L x,\qquad \Phi(X,T)=u(x,t),\qquad \Psi(X,T) =  \frac{D_{21}\sqrt{\kappa\beta}}{D_{22}^2L}v(x,t),
\end{equation}
and substituting into the \eqref{eq:brauns-marchetti-nrch} we then recover \eqref{BM1} which we rewrite as
\begin{subequations}\label{eq:nrch-nondim}
\begin{align}[left=\empheqlbrace]
    & \partial_t u = \Delta w, & x\in\Omega,\, t>0, \label{eq:nrch-nondim-u} \\
    & w = -\ep\Delta u + \ep^{-1}f(u) + \theta v, & x\in\Omega,\, t>0, \label{eq:nrch-nondim-w} \\
    & \tau\partial_t v = \Delta (v + \tau^{-1} u),& x\in\Omega,\, t>0 \label{eq:nrch-nondim-v}
\end{align}
\end{subequations}
where $\Omega:=[0,1]\times[0,\rho]$ and we define the non-dimensional parameters
\begin{equation}
    \ep := \frac{1}{L}\sqrt{\frac{\kappa}{\beta}},\qquad \tau := \frac{\sqrt{\kappa\beta}}{D_{22}L},\qquad \theta := \frac{D_{12}D_{21}}{D_{22}^2},\qquad \rho:=H/L.
\end{equation}
Our main results address the motion of the interface between regions where $u\approx 1$ and $u\approx -1$ in the sharp interface limit for which $\ep\ll 1$ is asymptotically small. Throughout the paper, we assume that the remaining problem parameters (i.e.\@ $\tau$, $\theta$, and $h$) are $O(1)$ with respect to $\ep\ll 1$.

Before stating our main results, we first fix the following notation. Let $\Lambda\subset\Omega$ be a two-dimensional subset with a smooth boundary, $\Gamma = \partial \Lambda$. For any $x\in\Gamma$, we denote by $\widehat{n}_\Gamma(x)$ the unit normal to $\Gamma$ at $x$ pointing towards the interior of $\Lambda$, and by $\widehat{\tau}_\Gamma(x)$ the unit tangent. Denote by $\kappa_\Gamma(x)$ the mean curvature of $\Gamma$ at $x$, with the sign chosen so that it is positive when $\Lambda$ is the unit ball. Next, we define the signed-distance from $\Gamma$ by
\begin{subequations}
\begin{equation}
    \dist(x,\Gamma) := \begin{cases} \min_{y\in\Gamma}\|x-y\|, & \text{for}\,x\in\Lambda, \\
    -\min_{y\in\Gamma}\|x-y\|, & \text{for}\,x\notin\Lambda.\end{cases}
\end{equation}
Finally, given a scalar or vector valued function $g(\cdot)$ that is continuous on $\Omega\setminus\Gamma$ we define
\begin{align}
    & \llbracket g\rrbracket_\Gamma(x) := \lim_{\substack{y\rightarrow x \\ y\in \Lambda}}g(y) - \lim_{\substack{y\rightarrow x \\ y\in \Omega\setminus\Lambda}}g(y),\qquad x\in\Gamma.
\end{align}
\end{subequations}
When the context is clear, we will often omit the explicit $x$ dependence in $\widehat{n}_\Lambda(x)$, $\kappa_\Gamma(x)$ and $\llbracket g\rrbracket_\Lambda(x)$. Finally, we denote by $|\Lambda|$ and $|\partial\Lambda|$ the area of $\Lambda$ and the length of its interface, respectively.

The following lemma is needed for the construction of inner solutions at the interface. Its proof is standard and can be approached using a phase-plane analysis.

\begin{lemma}\label{lem:heteroclinic}
    Let $f(z) = F'(z)$ where $F(\cdot)$ is a smooth balanced double-well potential with minima at $z=\pm 1$. Then there exists a unique heteroclinic $Q(\eta)$ solving
    \begin{equation}\label{eq:heteroclinic}
        \frac{d^2Q}{d\eta^2} - f(Q) = 0,\quad \eta\in\mathbb{R};\qquad Q(\eta)\rightarrow\pm 1,\quad \eta\rightarrow\pm\infty;\qquad Q(0)=0.
    \end{equation}
\end{lemma}

Our first main result is the following.

\begin{PR}\label{main-result-1}
    Fix $\Omega = [0,1]\times[0,\rho]$ and let $\Lambda_0$ be a smooth (not necessarily connected) two-dimensional subset of $\Omega$ with boundary $\Gamma_0$. Suppose that $v_0$ is a continuous and piecewise smooth function defined in $\Omega$. Let $\Lambda(t)$ be a time-dependent subset of $\Omega$ such that $\Lambda(0) = \Lambda_0$ and its boundary $\Gamma(t)=\partial\Lambda(t)$ has velocity $c_n$ in the direction of $\widehat{n}_{\Gamma(t)}$. Suppose that $w(x,t)$, $v(x,t)$, and $c_n$ satisfy the following modified Mullins-Sekerka (MS) system
    \begin{subequations}\label{eq:nrms}
    \begin{align}[left=\empheqlbrace]
        & \Delta w = 0, & x\in\Omega\setminus\Gamma(t),\, t>0, \label{eq:nrms-1} \\
        & \tau \partial_t v = \Delta v & x\in\Omega\setminus\Gamma(t),\, t>0, \label{eq:nrms-2} \\
        & w - \theta v = \gamma \kappa_{\Gamma(t)}, & x\in\Gamma(t),\, t>0, \label{eq:nrms-3} \\ 
        & \llbracket w \rrbracket_{\Gamma(t)} = \llbracket v \rrbracket_{\Gamma(t)} = 0, &x\in\Gamma(t),\, t>0, \label{eq:nrms-4} \\ 
        & \llbracket\nabla w\rrbracket_{\Gamma(t)}\cdot\widehat{n}_{\Gamma(t)} = -\llbracket \nabla v\rrbracket_{\Gamma(t)}\cdot\widehat{n}_{\Gamma(t)} = -2c_n, &x\in\Gamma(t),\, t>0, \label{eq:nrms-5}
    \end{align}
    \end{subequations}
    together with the initial condition $v(x,0)=v_0(x)$ and periodic boundary conditions on $\partial\Omega$. Above we have defined
    \begin{equation*}
        \gamma := \frac{1}{2}\int_{-\infty}^\infty \left|\frac{dQ}{d\eta}\right|^2 d\eta,
    \end{equation*}
    where $Q(\cdot)$ is the unique heteroclinic satisfying \eqref{eq:heteroclinic}. Then 
    \begin{equation}\label{eq:nrms-6}
        u_\ep(x,t) := Q\left(\tfrac{\dist(x,\Gamma(t))}{\ep}\right),\quad w_\ep(x,t) := w(x,t),\quad v_\ep(x,t) := v(x,t) - \frac{1}{\tau}Q\left(\tfrac{\dist(x,\Gamma(t))}{\ep}\right),
    \end{equation}
    is an asymptotic solution to the BM \eqref{eq:nrch-nondim} for $0<\ep\ll 1$.
\end{PR}

\begin{remark}\label{rem:boundary-conditions}
    For more general domains $\Omega$ with homogeneous Neumann boundary conditions (i.e.\@ $\partial_n u = \partial_n w= \partial_nv=0$ on $\Omega$) the above results are expected to remain true provided $\Lambda(t)$ is bounded away from the boundary $\partial\Omega$. We expect that the above results will remain true even when $\Gamma(t)$ intersects $\partial \Omega$, and moreover in such a case the intersection is perpendicular.
\end{remark}

Our primary use of Main Result \ref{main-result-1} will be to explicitly construct and study asymptotic approximations to the BM system \eqref{eq:nrch-nondim}. In \S \ref{sec:example-periodic}, we explore in detail the structure and stability of both stationary and traveling periodic wave-trains. To systematically characterize the linear stability of such solutions, we next formulate the appropriate eigenvalue problem.

Suppose that we can find a fixed $c_0\in\mathbb{R}^2$ and $\Lambda_0\subset\Omega$ with smooth boundary $\Gamma_0=\partial\Lambda_0$ such that $v(x,t)=v_0(x-c_0t)$ and $w(x,t)=w_0(x-c_0t)$ satisfy \eqref{eq:nrms} with 
\begin{equation}
    \Lambda(t) = \{x + c_0 t \in\mathbb{R}^2\,|\, x\in\Lambda_0\}.
\end{equation}
In particular, this includes both  stationary (if $c_0=0$) or uniformly traveling solutions. Changing to the moving reference frame $x\mapsto x-c_0t$ we find that $v_0(x)$ and $w_0(x)$ solve
\begin{subequations}\label{eq:nrms-moving}
\begin{align}[left=\empheqlbrace]
    & \Delta w = 0, & x\in\Omega\setminus\Gamma_0, \label{eq:nrms-moving-a}\\
    & \Delta v + \tau c_0\cdot\nabla v = 0 & x\in\Omega\setminus\Gamma_0, \label{eq:nrms-moving-b}\\
    & w - \theta v = \gamma \kappa_{\Gamma_0}, & x\in\Gamma_0, \label{eq:nrms-moving-c}\\ 
    & \llbracket w \rrbracket_{\Gamma_0} = \llbracket v \rrbracket_{\Gamma_0} = 0, &x\in\Gamma_0, \label{eq:nrms-moving-d}\\ 
    & \llbracket\nabla w\rrbracket_{\Gamma_0}\cdot\widehat{n}_{\Gamma_0} = -\llbracket \nabla v\rrbracket_{\Gamma(t)}\cdot\widehat{n}_{\Gamma(t)} = -2c_0\cdot\widehat{n}_{\Gamma_0}, &x\in\Gamma_0.\label{eq:nrms-moving-e}
\end{align}
\end{subequations}
The linear stability of such solutions with respect to arbitrary interface perturbations is determined by the following result.

\begin{PR}\label{main-result-2}
Let $\Lambda_0\subset\Omega$ be such that $w_0$, $v_0$, and $c_0$ solve \eqref{eq:nrms-moving}. Let $\zeta:\Gamma_0=\partial\Lambda_0\rightarrow\mathbb{R}$ be a smooth and arc-length parametrized function. Consider the following eigenvalue problem for $\varphi$, $\psi$, and $\lambda$
\begin{subequations}\label{eq:stability-problem-0}
\begin{align}[left=\empheqlbrace]
    & \Delta \varphi = 0, & x\in\Omega\setminus\Gamma_0, \label{eq:stability-problem-0-a}\\
    & \Delta \psi + \tau c_0 \cdot\nabla \psi - \tau\lambda \psi = 0, &  x\in\Omega\setminus\Gamma_0, \label{eq:stability-problem-0-b}\\
    & \llbracket \varphi\rrbracket_{\Gamma_0} = - \llbracket \psi\rrbracket_{\Gamma_0} = 2\zeta c_0\cdot\widehat{n}_{\Gamma_0}, & x\in\Gamma_0\label{eq:stability-problem-0-c} \\ 
    & \llbracket \nabla \varphi \rrbracket_{\Gamma_0}\cdot\widehat{n}_{\Gamma_0} = -2\lambda\zeta + 2\zeta' c_0\cdot\widehat{\tau}_{\Gamma_0}  - \zeta\widehat{n}_{\Gamma_0}\cdot\llbracket \mathrm{H}_{w_0}\rrbracket_{\Gamma_0}\widehat{n}_{\Gamma_0} + \zeta' \llbracket\nabla w_0\rrbracket_{\Gamma_0}\cdot\widehat{\tau}_{\Gamma_0}, & x\in\Gamma_0 \label{eq:stability-problem-0-d} \\
    & \llbracket \nabla \psi \rrbracket_{\Gamma_0}\cdot\widehat{n}_{\Gamma_0}  = 2\lambda\zeta - 2\zeta' c_0\cdot\widehat{\tau}_{\Gamma_0} - \zeta\widehat{n}_{\Gamma_0}\cdot\llbracket \mathrm{H}_{v_0}\rrbracket_{\Gamma_0}\widehat{n}_{\Gamma_0} + \zeta' \llbracket\nabla v_0\rrbracket_{\Gamma_0}\cdot\widehat{\tau}_{\Gamma_0}, & x\in\Gamma_0 \label{eq:stability-problem-0-e}\\
    & \lim_{\substack{x\rightarrow \Gamma_0}} \left[ \varphi-\theta\psi + \zeta\left( \nabla w_0 - \theta\nabla v_0 \right)\cdot \widehat{n}_{\Gamma_0}\right]= \gamma\left(\zeta'' + \kappa_{\Gamma_0}^2\zeta\right), & x\in\Gamma_0 \label{eq:stability-problem-0-f},
\end{align}
\end{subequations}
where $\mathrm{H}_{w_0}$ and $\mathrm{H}_{v_0}$ denote the Hessians of $w_0$ and $v_0$ respectively, and $\zeta'$ and $\zeta''$ denote the first and second derivatives of $\zeta$ with respect to arc-length. If the real part of $\lambda$ is negative for all $\zeta:\Gamma_0\rightarrow\mathbb{R}$ such that
\begin{equation}\label{eq:zero-delta-volume-constraint}
    \int_{\Gamma_0}\zeta(s)ds = 0,
\end{equation}
then the solution is linearly stable, and is linearly unstable otherwise.
\end{PR}

\begin{remark}
    The constraint \eqref{eq:zero-delta-volume-constraint} on $\zeta$ is needed to ensure that only volume-preserving interface perturbations are considered, and is in fact also needed for the solvability of \eqref{eq:stability-problem-0}.
\end{remark}

\begin{remark}
    Note that in domains with periodic or homogeneous Neumann boundary conditions, if $c_0=0$ then solutions $w_0$ and $v_0$ to \eqref{eq:nrms-moving} must be constant. As a consequence, in Main Result 2 the eigenvalue problem is significantly simplified since all derivatives of $w_0$ and $v_0$ must vanish.
\end{remark}

The remainder of the paper is organized as follows. In \S \ref{sec:main-results-deriations}, we derive Main Results \ref{main-result-1} and \ref{main-result-2}. In \S \ref{sec:example-periodic}, we study the dynamics of the modified MS system. We begin by comparing the properties of the classical and modified MS systems. Specifically, we show the modified MS system is area and mass preserving, but not, in general, length-shortening. We use a Lyapunov function to provide evidence that the dynamics of the modified MS are uninteresting when $\theta$ is positive. Moreover, we perform a detailed analysis of the structure of both stationary and traveling wave-trains. Finally, \S \ref{sec:discussion} summarizes our key findings and draws conclusions and suggestions for future work.

\section{Derivations of Main Results}\label{sec:main-results-deriations}

In this section, we derive Main Results \ref{main-result-1} and \ref{main-result-2} using a combination of matched asymptotic expansions and linear stability analysis. We first collect the needed geometric preliminaries in \S \ref{subsec:nrms-geo}. In \S \ref{subsec:nrms-derivation}, we derive Main Result \ref{main-result-1} and in \S \ref{subsec:nrms-stab-derivation}, we derive Main Result 2. 

\subsection{Geometric Preliminaries}\label{subsec:nrms-geo}

For each $t\geq 0$, let $X(s,t)$ be an arc-length parametrization of $\Gamma(t)=\partial\Lambda(t)$, i.e.
\begin{equation*}
    \Gamma(t) = \{X(s,t)\in\mathbb{R}^2\,|\, 0\leq X \leq |\Gamma(t)|\,,\, t>0\}.
\end{equation*}
We assume that $X(s,t)$ is at least twice differentiable in $s$ and once in $t\geq 0$. We then have
\begin{equation}\label{eq:frenet-serret}
    \frac{\partial X}{\partial s} = \widehat{\tau}_{\Gamma(t)},\qquad \frac{\partial \widehat{\tau}_{\Gamma(t)}}{\partial s} = \kappa_{\Gamma(t)} \widehat{n}_{\Gamma(t)},\qquad \frac{\partial \widehat{n}_{\Gamma(t)}}{\partial s} = -\kappa_{\Gamma(t)}\widehat{\tau}_{\Gamma(t)},
\end{equation}
where $\widehat{\tau}_{\Gamma(t)}(s)$ and $\widehat{n}_{\Gamma(t)}(s)$ are the unit normal and tangents to $\Gamma(t)$, and $\kappa_{\Gamma(t)}(s)$ is the curvature. Note that the latter two equations are the Frenet-Serret formulas. Finally, we remind the reader that we fix the parametrization orientation so that $\widehat{n}_{\Gamma(t)}$ is in the direction of $\Lambda(t)$ and $\Lambda(t)$ is to the left of $\widehat{\tau}_{\Gamma(t)}$.

We first collect expressions for spatial and temporal derivatives in terms of the interface-fitted coordinates $(s,\eta)$ given by
\begin{equation}\label{eq:rescaled-local-coordinates}
    x(s,t) = X(s,t) + \ep \eta \widehat{n}_{\Gamma(t)}(s).
\end{equation}
Note that this is a well-defined change of coordinates in an $O(\ep)$ region about $\Gamma(t)$ provided that $|\ep\eta|<\max_{0\leq s\leq |\Gamma(t)|}\{1/\kappa_{\Gamma(t)}(s)\}$. Assuming $\kappa_{\Gamma(T)}=O(1)$ for $0<\ep\ll 1$ then the interface-fitted coordinates are well defined for $|\eta|<O(1/\ep)$. Using \eqref{eq:frenet-serret}, we readily find that the corresponding metric tensor is given by
\begin{equation*}
    g = \begin{pmatrix} \frac{\partial x}{\partial s}\cdot\frac{\partial x}{\partial s} & \frac{\partial x}{\partial s}\cdot\frac{\partial x}{\partial \eta} \\ \frac{\partial x}{\partial \eta}\cdot\frac{\partial x}{\partial s} & \frac{\partial x}{\partial \eta}\cdot\frac{\partial x}{\partial \eta}
    \end{pmatrix}  =  \begin{pmatrix}
        (1-\ep\kappa_{\Gamma(t)}\eta)^2 & 0 \\ 0 & \ep^2
    \end{pmatrix}.
\end{equation*}
It follows that the gradient and Laplacian in the interface-fitted coordinates are, respectively, given by
\begin{subequations}
\begin{align}
    \nabla \varphi & = \frac{1}{1-\ep\kappa_{\Gamma(t)}\eta}\widehat{\tau}_{\Gamma(t)}\frac{\partial \varphi}{\partial s} + \frac{1}{\ep}\widehat{n}_{\Gamma(t)}\frac{\partial \varphi}{\partial\eta}, \label{eq:interface-fitted-gradient}\\
    \Delta\varphi & = \frac{1}{\ep^2}\frac{\partial^2\varphi}{\partial \eta^2} - \frac{1}{\ep}\frac{\kappa_{\Gamma(t)}}{1-\ep\kappa_{\Gamma(t)}\eta}\frac{\partial\varphi}{\partial\eta} + \frac{1}{1-\ep\kappa_{\Gamma(t)}\eta}\frac{\partial}{\partial s}\left(\frac{1}{1-\ep\kappa_{\Gamma(t)}\eta}\frac{\partial\varphi}{\partial s}\right), \label{eq:interface-fitted-laplacian-0}
\end{align}
for any smooth function $\varphi$. In particular, if $\varphi$ and its $s$- and $\eta$-derivatives are $O(1)$ for $0<\ep\ll 1$, then
\begin{equation}\label{eq:interface-fitted-laplacian}
    \Delta\varphi = \frac{1}{\ep^2}\frac{\partial^2\varphi}{\partial\eta^2} - \frac{\kappa_{\Gamma(t)}}{\ep}\frac{\partial\varphi}{\partial\eta} + O(1).
\end{equation}
\end{subequations}
Finally, given the time-evolution $\partial_t X(s,t) = c_n\widehat{n}_{\Gamma(t)}$, we deduce that
\begin{equation}\label{eq:interface-fitted-time-derivative}
    \frac{d}{dt}\varphi(x-X(s,t)) = \frac{\partial\varphi}{\partial t} - \frac{\partial X}{\partial t}\cdot\nabla \varphi  = \frac{\partial\varphi}{\partial t} - \frac{1}{\ep}c_n\frac{\partial\varphi}{\partial \eta}.
\end{equation}

We next collect key results for how small perturbations to the interface affect key geometric quantities. Suppose $\delta\ll 1$ and let
\begin{equation}\label{eq:Gamma_0_Gamma_delta}
    \Gamma_0 = \{ X_{0}(s) \,|\, 0\leq s<|\Gamma_0|\}\quad\text{and}\quad \Gamma_\delta(t) = \{ X_{\delta}(s,t) \,|\, 0\leq s <|\Gamma_0|\},
\end{equation}
were $X_0:[0,|\Gamma_0|]$ is an arc-length parametrization of $\Gamma_0$ and 
\begin{equation}
    X_\delta(s,t) := X_0(s) + c_0 t + \delta \zeta(s)e^{\lambda t} \widehat{n}_{\Gamma_0}(s),
\end{equation}
where $c_0\in\mathbb{R}^2$, $\delta\in\mathbb{R}$ is a small parameter and $\zeta:[0,|\Gamma_0|]\rightarrow \mathbb{R}$ is a smooth function. The additional term $e^{\lambda t}$ is included to aid in the derivation of Main Result \ref{main-result-2} found in \S\ref{subsec:nrms-stab-derivation} below.

Note that $X_\delta(s,t)$ is in general not an arc-length parametrization of $\Gamma_\delta(t)$. Using \eqref{eq:frenet-serret}, we then calculate
\begin{align*}
    & X_{\delta}'(s,t) = \left(1 - \delta \kappa_{\Gamma_0}(s) \zeta(s)e^{\lambda t}\right)\widehat{\tau}_{\Gamma_0}(s) + \delta\zeta'(s)e^{\lambda t}\widehat{n}_{\Gamma_0}(s), \\
    & X_{\delta}''(s,t) = -\delta\left(\kappa_{\Gamma_0}'(s)\zeta(s)+2\kappa_{\Gamma_0}(s)\zeta'(s)\right)e^{\lambda t}\widehat{\tau}_{\Gamma_0}(s) + \left(\kappa + \delta\left(\zeta''(s) - \kappa_{\Gamma_0}(s)^2\zeta(s)\right)e^{\lambda t} \right)\widehat{n}_{\Gamma_0}(s),
\end{align*}
where $'$ denotes differentiation with respect to $s$. The arc-length along $\Gamma_\delta$ is then given by
\begin{equation*}
    \sigma(s,t) = \int_0^s \|X_{\delta}'(\tilde{s},t)\|d\tilde{s} = \int_0^s \sqrt{1 - 2\delta \kappa_{\Gamma_0}(\tilde{s})\zeta(\tilde{s})e^{\lambda t} + O(\delta^2)}d\tilde{s}.
\end{equation*}
Then from 
\begin{equation*}
    \widehat{\tau}_{\Gamma_\delta(t)}(s) = \frac{X_{\delta}'(s,t)}{\sigma'(s)},\qquad \frac{d}{d\sigma}\widehat{\tau}_{\Gamma_\delta(t)}(s) = \frac{1}{(\sigma'(s))^2}X_{\delta}''(s,t)  - \frac{\sigma''(s)}{(\sigma'(s))^3}X_{\delta}'(s,t),
\end{equation*}
and the Frenet-Serret formulas \eqref{eq:frenet-serret} we find
\begin{subequations}\label{eq:perturbed-geometric-quantities}
\begin{align}
    \widehat{\tau}_{\Gamma_\delta(t)}(s) & = \widehat{\tau}_{\Gamma_0}(s) + \delta\zeta'(s)e^{\lambda t}\widehat{n}_{\Gamma_0}(s) + O(\delta^2), \\
    \widehat{n}_{\Gamma_\delta(t)}(s) & = \widehat{n}_{\Gamma_0}(s) - \delta \zeta'(s)e^{\lambda t}\widehat{\tau}_{\Gamma_0}(s) + O(\delta^2), \\
    \kappa_{\Gamma_\delta(t)}(s) & =  \kappa_{\Gamma_0}(s) + \delta \left(\zeta''(s) + \kappa_{\Gamma_0}(s)^2\zeta(s)  \right)e^{\lambda t} + O(\delta^2).
\end{align}
Finally, the normal speed of $\Gamma_\delta(t)$ is given by
\begin{equation}
    c_n  = \frac{\partial X_\delta(s,t)}{\partial t}\cdot\widehat{n}_{\Gamma_\delta(t)}(s) = c_0\cdot\widehat{n}_{\Gamma_0}(s) + \delta e^{\lambda t}\left( \lambda\zeta(s) - \zeta'(s)c_0\cdot\widehat{\tau}_{\Gamma_0}(s)   \right) + O(\delta^2).
\end{equation}
\end{subequations}

\subsection{Derivation of Main Result 1}\label{subsec:nrms-derivation}

We now derive Main Result \ref{main-result-1} using the method of matched asymptotic expansions. To start, we replace $v\mapsto v - \tau^{-1}u$ so \eqref{eq:nrch-nondim} becomes
\begin{subequations}\label{eq:nrch-nondim_}
\begin{align}[left=\empheqlbrace]
    & \partial_t u = \Delta w, & x\in\Omega,\, t>0, \label{eq:nrch-nondim-u_} \\
    & w = -\ep\Delta u + \ep^{-1}f(u) + \theta \left(v-\tau^{-1}u\right), & x\in\Omega,\, t>0, \label{eq:nrch-nondim-w_} \\
    & \tau\partial_t v = \Delta (v + w),& x\in\Omega,\, t>0. \label{eq:nrch-nondim-v_}
\end{align}
\end{subequations}

We first seek an outer solution that is valid for values of $x$ that are sufficiently far from $\Gamma(t)$ in the sense that $\dist(x,\Gamma(t))\gg \ep$. We seek a regular asymptotic expansion of the form
\begin{align*}[left=\empheqlbrace]
    & u(x,t) = u_0(x,t) + \ep u_1(x,t) + O(\ep^2),\\
    & w(x,t) = w_0(x,t) + O(\ep),\\
    & v(x,t) = v_0(x,t) + O(\ep).
\end{align*}
Substituting into \eqref{eq:nrch-nondim-u_} yields the leading order equation $f(u_0)=0$ from which we deduce that
\begin{equation}
    u_0(x,t) = \begin{cases} 1, & x\in\Lambda(t), \\ -1, & x\in\Omega\setminus\Lambda(t).\end{cases}
\end{equation}
Collecting next the $O(1)$ equations in \eqref{eq:nrch-nondim_} we obtain
\begin{equation*}
    \partial_t u_0 = \Delta w_0,\quad w_0 = f_u(u_0)u_1 + \theta \left(v_0-\tau^{-1}u\right),\quad \tau\partial_t v = \Delta(v_0 + w_0),
\end{equation*}
from which we deduce
\begin{equation}
    \Delta w_0 = 0\quad\text{and}\quad \tau \partial_t v_0 = \Delta v_0.
\end{equation}
These two equations are posed on $\Omega\setminus\Gamma(t)$ with periodic boundary conditions being imposed on $\partial\Omega$. To determine $\Gamma(t)$ we must formulate an appropriate inner problem.

Let $(s,\eta)$ be the scaled interface-fitted coordinates introduced in \eqref{eq:rescaled-local-coordinates} and define $U$, $W$, and $V$ by
\begin{equation*}
    u(x,t) = U(s,\eta,t),\quad w(x,t) = W(s,\eta,t),\quad v(x,t) = V(s,\eta,t).
\end{equation*}
Substituting into \eqref{eq:nrch-nondim_} and using the expression \eqref{eq:interface-fitted-laplacian} for the Laplacian we obtain
\begin{subequations}\label{eq:inner-problem-1}
\begin{align}[left=\empheqlbrace]
    & \frac{\partial U}{\partial t} - \frac{1}{\ep}c_n\frac{\partial U}{\partial\eta} = \frac{1}{\ep^2}\frac{\partial^2 W}{\partial\eta^2} - \frac{\kappa_{\Gamma(t)}}{\ep}\frac{\partial W}{\partial\eta} + \hot, \label{eq:inner-problem-1-a}\\
    & W = -\frac{1}{\ep}\frac{\partial^2 U}{\partial\eta^2} + \kappa_{\Gamma(t)}\frac{\partial U}{\partial \eta} + \frac{1}{\ep}f(U) + \theta \left(V-\tau^{-1}U\right) + \hot, \label{eq:inner-problem-1-b}\\
    & \tau\frac{\partial V}{\partial t} - \frac{1}{\ep}\tau c_n\frac{\partial V}{\partial\eta} = \frac{1}{\ep^2}\frac{\partial^2}{\partial\eta^2}\left(V + W\right) - \frac{\kappa_{\Gamma(t)}}{\ep}\frac{\partial}{\partial\eta}\left(V + W\right) + \hot, \label{eq:inner-problem-1-c}
\end{align}
\end{subequations}
where $\hot$ denotes higher-order-terms. We seek an asymptotic approximation of the form
\begin{align*}[left=\empheqlbrace]
    & U(\eta,s,t) = U_0(\eta,s,t) + \ep U_1(\eta,s,t) + O(\ep^2), \\
    & W(\eta,s,t) = W_0(\eta,s,t) + \ep W_1(\eta,s,t) + O(\ep^2), \\
    & V(\eta,s,t) = V_0(\eta,s,t) + \ep V_1(\eta,s,t) + O(\ep^2).
\end{align*}
Substituting into \eqref{eq:inner-problem-1} and collecting different order in $\ep$ we obtain a sequence of inner problems. Specifically, from \eqref{eq:inner-problem-1-b} we obtain the $O(\ep^{-1})$ order problem
\begin{equation*}
    -\frac{\partial^2 U_0}{\partial\eta^2} + f(U_0) = 0.
\end{equation*}
The far-field behavior of $U_0$ must coincide with the outer solution so $U_0\rightarrow \pm 1$ as $\eta\rightarrow \pm\infty$. This implies that $U_0(\eta,s,t) = Q(\eta)$ where $Q(\eta)$ is the unique heteroclinic in Lemma \ref{lem:heteroclinic}. Note that we have implicitly fixed the interface $\Gamma(t)$ to coincide with the level set where $u(x,t)=0$.

Next from \eqref{eq:inner-problem-1-a} and \eqref{eq:inner-problem-1-c} we obtain the $O(\ep^{-2})$ problems
\begin{equation*}
    \frac{\partial^2 W_0}{\partial\eta^2} =  0\quad\text{and}\quad \frac{\partial^2}{\partial\eta^2}\left(V_0+W_0\right) = 0.
\end{equation*}
Both $W_0$ and $V_0+W_0$ must be linear functions of $\eta$. However, since $W_0$ and $V_0$ must remain bounded as $\eta\rightarrow\pm\infty$, we deduce that in fact both of these quantities are constants in $\eta$, and therefore, by matching with the limiting values of $w_0$ and $v_0$ as the interface is approached, we deduce
\begin{equation}\label{eq:W0-V0-1}
    W_0=w_0|_{\Gamma(t)}\quad\text{and}\quad V_0 = v_0|_{\Gamma(t)}.
\end{equation}
Note in particular that this implies that both $w_0$ and $v_0$ must be continuous across the interface.

Next from \eqref{eq:inner-problem-1-b} which we obtain the $O(1)$ problem 
\begin{equation}\label{eq:inner-problem-2}
    W_0 - \theta V_0 + \tau^{-1}\theta Q = -\frac{\partial^2 U_1}{\partial\eta^2} + f_u(Q)U_1 + \kappa_{\Gamma(t)} \frac{d Q}{d\eta},
\end{equation}
where $f_u(\cdot)$ denotes the derivative of $f$ with respect to $u$, and for which we impose that $U_1\rightarrow 0$ as $\eta\rightarrow\pm\infty$. Differentiating \eqref{eq:heteroclinic}, we find that $dQ/d\eta$ satisfies the homogeneous equation
$$
-\frac{d^2}{d\eta^2}\left(\frac{dQ}{d\eta}\right) + f_u(Q)\frac{dQ}{d\eta} = 0.
$$
Therefore, multiplying \eqref{eq:inner-problem-2} by $\frac{dQ}{d\eta}$ and integrating, we get the solvability condition
\begin{equation*}
    w_0|_{\Gamma(t)}-\theta v_0|_{\Gamma(t)} = \frac{\kappa_{\Gamma(t)}\int_{-\infty}^\infty \left|\frac{dQ}{d\eta}\right|^2d\eta - \frac{\theta}{\tau}\int_{-\infty}^\infty Q\frac{dQ}{d\eta}d\eta}{\int_{-\infty}^\infty \frac{dQ}{d\eta}d\eta} = \gamma\kappa_{\Gamma(t)},
\end{equation*}
where $\gamma := \frac{1}{2}\int_{-\infty}^\infty |\frac{dQ}{d\eta}|^2d\eta$.

Finally, from \eqref{eq:inner-problem-1-a} and \eqref{eq:inner-problem-1-c} we obtain the $O(\ep^{-1})$ problems
\begin{subequations}
\begin{align}[left=\empheqlbrace]
    & -c_n \frac{dQ}{d\eta} = \frac{\partial^2 W_1}{\partial\eta^2}, \label{eq:inner-problem-3-a}\\
    & -\tau c_n \frac{\partial V_0}{\partial\eta} = \frac{\partial^2}{\partial\eta^2}\left(V_1+W_1\right) - \kappa_{\Gamma(t)}\frac{\partial}{\partial\eta}\left(V_0+W_0\right). \label{eq:inner-problem-3-b}
\end{align}
\end{subequations}
Note that
\begin{align*}
    &\lim_{\substack{x\rightarrow \Gamma(t)\\ x\in\Lambda(t)}} \nabla w_0(x,t)\cdot\widehat{n}_{\Gamma(t)} = \lim_{\eta\rightarrow+\infty} \frac{\partial W_1}{\partial\eta}, & \lim_{\substack{x\rightarrow \Gamma(t)\\ x\in\Omega\setminus\Lambda(t)}} \nabla w_0(x,t)\cdot\widehat{n}_{\Gamma(t)} = \lim_{\eta\rightarrow-\infty} \frac{\partial W_1}{\partial\eta}, \\
    &\lim_{\substack{x\rightarrow \Gamma(t)\\ x\in\Lambda(t)}} \nabla v_0(x,t)\cdot\widehat{n}_{\Gamma(t)} = \lim_{\eta\rightarrow+\infty} \frac{\partial V_1}{\partial\eta}, & \lim_{\substack{x\rightarrow \Gamma(t)\\ x\in\Omega\setminus\Lambda(t)}} \nabla v_0(x,t)\cdot\widehat{n}_{\Gamma(t)} = \lim_{\eta\rightarrow-\infty} \frac{\partial V_1}{\partial\eta}.
\end{align*}
Therefore, integrating \eqref{eq:inner-problem-3-a} and \eqref{eq:inner-problem-3-b} respectively gives
\begin{equation*}
    \llbracket \nabla w_0\rrbracket_{\Gamma(t)}\cdot\widehat{n}_{\Gamma(t)} = -2 c_n,\qquad \llbracket \nabla v_0\rrbracket_{\Gamma(t)}\cdot\widehat{n}_{\Gamma(t)} =  - \llbracket \nabla w_0\rrbracket_{\Gamma(t)}\cdot\widehat{n}_{\Gamma(t)} = 2 c_n.
\end{equation*}

We have thus shown that $w_0$ and $v_0$ must satisfy the modified MS equations \eqref{eq:nrms}. To deduce \eqref{eq:nrms-6} note that $u_\ep(x,t)$ is the composite solution and then recall that we had replaced $v\mapsto v- \tau^{-1}u$.

\subsection{Derivation of Main Result 2}\label{subsec:nrms-stab-derivation}

To derive Main Result \ref{main-result-2} we consider the modified MS dynamics \eqref{eq:nrms} in the moving reference frame $x\mapsto x - c_0 t$ 
\begin{subequations}\label{eq:stability-problem-1}
\begin{align}[left=\empheqlbrace]
    & \Delta w_\delta = 0, & x\in\Omega\setminus\Gamma_\delta(t),\, t>0, \label{eq:stability-problem-1-a} \\
    & \tau\partial_t v_\delta = \Delta v_\delta + \tau c_0 \cdot \nabla v_\delta, & x\in\Omega\setminus\Gamma_\delta(t),\, t>0, \label{eq:stability-problem-1-b} \\
    & w_\delta - \theta v_\delta = \gamma \kappa_{\Gamma_\delta(t)}, & x\in\Gamma_\delta(t),\, t>0, \label{eq:stability-problem-1-c}\\ 
    & \llbracket w_\delta \rrbracket_{\Gamma_\delta(t)} = \llbracket v_\delta \rrbracket_{\Gamma_\delta(t)} = 0, &t>0, \label{eq:stability-problem-1-d}\\
    & \llbracket\nabla w_\delta\rrbracket_{\Gamma_\delta(t)}\cdot\widehat{n}_{\Gamma_\delta(t)} = -\llbracket \nabla v\rrbracket_{\Gamma_\delta(t)}\cdot\widehat{n}_{\Gamma_\delta(t)} = -2c_n, &t>0, \label{eq:stability-problem-1-e}
\end{align}
\end{subequations}
with perturbed initial conditions
\begin{equation*}
    w_\delta(x,0) = w_0(x) + \delta \varphi(x),\quad v_\delta(x,0) = v_0(x) + \delta \psi(x),
\end{equation*}
and
\begin{equation*}
    \Gamma_\delta(0)=\partial\Lambda_\delta(0) = \{X_0(s) + \delta \zeta(s)\widehat{n}_{\Gamma_0}\,|\, 0\leq s\leq |\Gamma_0|\}
\end{equation*}
where $X_0(s)$ is an arc-length parametrization of $\Gamma_0$ and $\zeta:[0,|\Gamma_0|]\rightarrow\mathbb{R}$ is an arbitrary smooth (and periodic, if $\Gamma_0$ is closed) function. To determine whether $(w_\delta(x,t), v_\delta(x,t),\Lambda_\delta(t))$ relax back to $(w_0(x),v_0(x),\Lambda_0)$ we consider the linearized problem by seeking
\begin{subequations}\label{eq:stability-problem-2-3}
\begin{equation}\label{eq:stability-problem-2}
    w_\delta(x,t) = w_0(x) + \delta e^{\lambda t}\varphi(x),\qquad v_\delta(x,t) = v_0(x) + \delta e^{\lambda t}\psi(x),
\end{equation}
and
\begin{equation}\label{eq:stability-problem-3}
    \Gamma_\delta(t)=\partial\Lambda_\delta(t) = \{X_\delta(s,t)\,|\, 0\leq s\leq |\Gamma_0|\},\quad X_\delta(s,t) := X_0(s) + \delta \zeta(s)e^{\lambda t}\widehat{n}_{\Gamma_0}.
\end{equation}
\end{subequations}
Since $w_0$ and $v_0$ satisfy \eqref{eq:nrms-moving}, substituting \eqref{eq:stability-problem-2-3} into \eqref{eq:stability-problem-1-a} and \eqref{eq:stability-problem-1-b} immediately gives \eqref{eq:stability-problem-0-a} and \eqref{eq:stability-problem-0-b} respectively.

To derive the remaining equations in \eqref{eq:stability-problem-0} we need to approximate all quantities defined on $\Gamma_\delta(t)$ to ones defined on $\Gamma_0$. In \eqref{eq:perturbed-geometric-quantities} we have already established expressions for $\widehat{\tau}_{\Gamma_\delta(t)}$, $\widehat{n}_{\Gamma_\delta(t)}$, $\kappa_{\Gamma_\delta(t)}$, and $c_n$. It therefore remains only to determine how the jump operator $\llbracket\cdot\rrbracket_{\Gamma_\delta(t)}$ must be modified.

For a given  $0<\phi<\pi$ we define
$$
q_\delta(\phi) = \widehat{\tau}_{\Gamma_\delta(t)}(s)\cos\phi + \widehat{n}_{\Gamma_\delta(t)}(s)\sin\phi = \widehat{\tau}_{\Gamma_0}(s)\cos\phi + \widehat{n}_{\Gamma_0}(s)\sin\phi + O(\delta) =: q_0(\phi) + O(\delta).
$$
Then for any \textit{nonzero} $\nu$ such that $\nu\ll\delta$ we calculate
\begin{gather*}
    f(X_{\delta}(s,t) + \nu q_\delta(\phi)) = f(X_{0}(s) + \nu q_0) + \delta\zeta(s)e^{\lambda t}\widehat{n}_{\Gamma_0}(s)\cdot\nabla f(X_{0}(s) +\nu q_0) + O(\delta^2), \\
    \nabla f(X_{\delta}(s,t) + \nu q_\delta(\phi)) = \nabla f(X_{0}(s) + \nu q_0) + \delta\zeta(s)e^{\lambda t} \mathrm{H}_f (X_{0}(s) +\nu q_0)\widehat{n}_{\Gamma_0}(s) + O(\delta^2).
\end{gather*}
Taking the limits $\nu\rightarrow 0^\pm$ then yields
\begin{subequations}
\begin{gather}
    \llbracket f\rrbracket_{\Gamma_\delta(t)} =  \llbracket f\rrbracket_{\Gamma_0} + \delta\zeta e^{\lambda t}\llbracket\nabla f\rrbracket_{\Gamma_0}\cdot \widehat{n}_{\Gamma_0} + O(\delta^2), \label{eq:perturbed-func}\\
    \llbracket\nabla f\rrbracket_{\Gamma_\delta(t)}\cdot\widehat{n}_{\Gamma_\delta(t)} = \llbracket\nabla f\rrbracket_{\Gamma_0}\cdot\widehat{n}_{0} + \delta e^{\lambda t}\left(\zeta \widehat{n}_{\Gamma_0}\cdot \llbracket\mathrm{H}_f\rrbracket_{\Gamma_0}\widehat{n}_{\Gamma_0} - \zeta' \llbracket\nabla f\rrbracket_{\Gamma_0}\cdot\widehat{\tau}_{\Gamma_0}   \right) + O(\delta^2),\label{eq:perturbed-grad}
\end{gather}
\end{subequations}
Using \eqref{eq:perturbed-func} together with \eqref{eq:nrms-moving-d}, \eqref{eq:nrms-moving-e} and \eqref{eq:stability-problem-2} in \eqref{eq:stability-problem-1-d} establishes \eqref{eq:stability-problem-0-c}. Similarly, using \eqref{eq:perturbed-grad} together with \eqref{eq:nrms-moving-e} and \eqref{eq:stability-problem-2} in \eqref{eq:stability-problem-1-e} establishes \eqref{eq:stability-problem-0-d} and \eqref{eq:stability-problem-0-e}.

Finally, to show \eqref{eq:stability-problem-0-f}, we evaluate $w_\delta - \theta v_\delta$ at $x=X_\delta(s,t)+\nu q_\delta(\varphi)$ for $\nu\ll\delta$
\begin{equation*}
    w_\delta -\theta v_\delta = (w_0-\theta v_0)|_{x=X_0+\nu q_0} + \delta e^{\lambda t}\left(\varphi - \theta \psi + \zeta \left(\nabla w_0 - \theta \nabla v_0\right)\cdot\widehat{n}_{\Gamma_0}\right)|_{x=X_0+\nu q_0} + O(\delta^2).
\end{equation*}
Using \eqref{eq:stability-problem-0-c}--\eqref{eq:stability-problem-0-e} implies that the $O(\delta)$ term above satisfies
\begin{equation*}
    \llbracket \varphi - \theta \psi + \zeta \left(\nabla w_0 - \theta \nabla v_0\right)\cdot\widehat{n}_{\Gamma_0} \rrbracket_{\Gamma_0} = 0,
\end{equation*}
so that it is continuous. Therefore, taking the limit $\nu\rightarrow 0$ and using \eqref{eq:perturbed-geometric-quantities} for $\kappa_{\Gamma_\delta(t)}$ establishes \eqref{eq:stability-problem-0-f}.

\section{Dynamics of the Modified Mullins-Sekerka System}\label{sec:example-periodic}

In this section, we consider the dynamics exhibited by the modified MS system. In \S\ref{subsec:nrms-properties}, we first derive some key properties of the modified MS system \eqref{eq:nrms}, showing in particular that the dynamics for $\theta>0$ are effectively governed by a gradient flow dynamics and therefore can't support rich spatiotemporal behavior. For $\theta<0$, the behavior can however be more intricate and we illustrate this in \S\ref{subsec:example-periodic-1}-\ref{subsec:example-periodic-3} by considering in detail the structure and dynamics of periodic wave-trains. Specifically, using Main Result \ref{main-result-1}, we demonstrate that for $-1<\theta<0$ only stationary periodic wave-trains are possible, but for $\theta<-1$ the system exhibits traveling periodic wave-trains whose speed is governed by a remarkably simple transcendental equation. Then, using Main Result \ref{main-result-2} together with a winding-number argument, we demonstrate that the stationary periodic wave-train solutions are linearly stable only for $\theta>-1$ and become unstable with respect to a translational mode otherwise. On the other hand, traveling periodic wave-trains are linearly stable for only a finite range of $\theta<-1$ values, beyond which it becomes unstable with respect to non-translational modes of non-zero transverse wavelength. In \S\ref{subsec:example-periodic-4}, we numerically validate the predictions of the modified MS system by comparing with full numerical simulations of the BM model \eqref{BM} using FlexPDE7 \cite{flexpde}. In particular, we numerically observe that the transverse instability of traveling periodic wave-trains leads to a new seemingly stable wave-train with undulating interfaces.

\subsection{Properties of the modified MS System}\label{subsec:nrms-properties}

The classical MS system is known to be a area preserving and length-shortening flow. In the case of the modified MS system \eqref{eq:nrms} the situation is more complicated, particularly for $\theta<0$. We first consider some of the similarities with the classical MS system. By the divergence theorem and \eqref{eq:nrms-5}, we see
\begin{equation*}
    \int_{\Gamma(t)} c_nds = -\frac{1}{2}\int_{\Gamma(t)} \llbracket \nabla w\rrbracket_{\Gamma(t)}\cdot\widehat{n}_{\Gamma(t)}ds = -\frac{1}{2}\int_{\Omega\setminus\Gamma(t)}\Delta w dx = 0.
\end{equation*}
Next, using the Reynolds-Leibniz and divergence theorems, we calculate
\begin{align*}
    \frac{d}{dt}\int_{\Omega\setminus\Gamma(t)} v(x,t)dx & = \frac{d}{dt}\int_{\Lambda(t)}v(x,t)dx + \frac{d}{dt}\int_{\Omega\setminus\Lambda(t)}v(x,t)dx \\
    & = -\frac{1}{\tau}\int_{\Gamma(t)}\llbracket \nabla v\rrbracket_{\Gamma(t)}\cdot\widehat{n}_{\Gamma(t)}ds- \int_{\Gamma(t)}c_n\llbracket v\rrbracket_{\Gamma(t)} ds \\
    & = -\frac{2}{\tau}\int_{\Gamma(t)}c_nds = 0.
\end{align*}
On the other hand, by directly applying the Reynolds-Leibniz theorem, we immediately get
\begin{align*}
    \frac{d}{dt}|\Lambda(t)| = \frac{d}{dt}\int_{\Lambda(t)}dx = -\int_{\Gamma(t)}c_nds = 0.
\end{align*}
In summary, we have found that
\begin{equation}
    \frac{d}{dt}|\Lambda(t)| = 0,\qquad \frac{d}{dt}\int_\Omega v(x,t)dx = 0,
\end{equation}
so, the dynamics are area and mass preserving. 

Next we show that the modified MS dynamics are in general not length-shortening but does satisfy a related identity. Indeed, we calculate
\begin{align*}
    \frac{d}{dt}|\Gamma(t)| & = -\int_{\Gamma(t)}c_n\kappa_{\Gamma(t)}ds = -\frac{1}{\gamma}\int_{\Gamma(t)} c_n(w-\theta v)ds \\
    & = \frac{1}{2\gamma}\int_{\Gamma(t)}\left(\llbracket w\nabla w\rrbracket_{\Gamma(t)}\cdot\widehat{n}_{\Gamma(t)} + \theta \llbracket v\nabla v\rrbracket_{\Gamma(t)}\cdot\widehat{n}_{\Gamma(t)} \right)ds \\
    & = -\frac{1}{2\gamma}\int_{\Omega\setminus\Gamma(t)}\left(\|\nabla w\|^2 + \theta\|\nabla v\|^2 + \theta v\Delta v\right)dx.
\end{align*}
On the other hand,
\begin{equation*}
    \int_{\Omega\setminus\Gamma(t)}v\Delta v = \tau\int_{\Omega\setminus\Gamma(t)}v\frac{\partial v}{\partial t}dx = \frac{\tau}{2}\int_{\Omega\setminus\Gamma(t)}\frac{\partial (v^2)}{\partial t}dx = \frac{\tau}{2}\frac{d}{dt}\int_{\Omega\setminus\Gamma(t)}v^2dx,
\end{equation*}
so that 
\begin{equation}\label{eq:lyapunov-nrms}
    \frac{d}{dt}\left(|\Gamma(t)| + \frac{\tau\theta}{4\gamma}\int_\Omega v^2 dx   \right) = -\frac{1}{2\gamma}\int_{\Omega\setminus\Gamma(t)}\left(\|\nabla w\|^2 + \theta \|\nabla v\|^2\right)dx.
\end{equation}
If $\theta>0$, then this defines a Lyapunov function which implies that the dynamics eventually settle to a stationary solution. In fact, more can be said by noting that with the rescaling $v= \widetilde{v}/\tau\sqrt{\theta}$ the BM model \eqref{BM} becomes
\begin{equation*}
    \PD{u}{t}=\Delta \paren{-\ep \Delta u+\frac{1}{\ep}f(u)+\sqrt{\frac{\theta}{\tau}} \widetilde{v}},\qquad 
\tau\PD{\widetilde{v}}{t}=\Delta\left(\widetilde{v}+\sqrt{\frac{\theta}{\tau}}u\right)
\end{equation*}
which is the $H^{-1}(\Omega)\times H^{-1}(\Omega)$ gradient flow of the energy
\begin{equation}
    \widetilde{\mathcal{E}}[u,\widetilde{v}] = \int_\Omega\left(\frac{\ep}{2}\|\nabla u\|^2 + \frac{1}{\ep}F(u) + \frac{1}{2\tau}(\widetilde{v}+\sqrt{\theta}u)^2 - \frac{\theta}{2\tau}u^2\right) dx.
\end{equation}
Thus for $\theta>0$ we expect the dynamics of the BM model to be qualitatively similar to the classical CH model \eqref{CH}. 

The above discussion suggests dynamics that are distinct from the classical MS system are possible only for $\theta<0$. In the remainder of this section we present some results suggesting that in fact $\theta<-1$ is needed. In \S\ref{subsec:example-periodic-1}-\S\ref{subsec:example-periodic-3} we use Main Results \ref{main-result-1} and \ref{main-result-2} to perform a detailed analysis of periodic wave-trains, explicitly showing that $\theta<-1$ is needed for nontrivial temporal dynamics, mainly traveling waves. Before we delve into this discussion, we note here that this threshold naturally arises by considering the well-posedness of the modified MS system \eqref{eq:nrms} when $\tau=0$. 
Indeed, when $\tau=0$ we may simplify \eqref{eq:nrms} by setting $v=-w$ to get the system
\begin{subequations}
\begin{align}[left=\empheqlbrace]
    & \Delta w = 0, & x\in\Omega\setminus\Gamma(t),\, t>0, \\
    & (1+\theta)w = \gamma \kappa_{\Gamma(t)}, & x\in\Gamma(t),\, t>0,\\ 
    & \llbracket w \rrbracket_{\Gamma(t)} = 0, &x\in\Gamma(t),\, t>0,\\ 
    & \llbracket\nabla w\rrbracket_{\Gamma(t)}\cdot\widehat{n}_{\Gamma(t)} = -2c_n, &x\in\Gamma(t),\, t>0.
\end{align}
\end{subequations}
The above is nothing other than the classical MS system when $(1+\theta)>0$. It is well-known that the principal part dynamics of the MS model is third-order diffusion \cite{xinfu1996existance}. If $(1+\theta)<0$, the principal evolution will be that of a third-order backward diffusion, making the equations ill-posed. When $\tau>0$, this limiting ill-posedness can be interpreted as giving rise to instabilities.

In the remainder of this section, we use Main Results \ref{main-result-1} and \ref{main-result-2} to explore the structure and dynamics of stationary and traveling periodic wave-train solutions. Specifically, we analyze in detail the structure and linear stability of solutions to \eqref{eq:nrms} with
\begin{equation}
    \Lambda(t) : = \bigcup_{n=0}^{N-1}[x_{2n}+c_0t,x_{2n+1}+c_0t]\times[0,\rho],
\end{equation}
where $N\geq 1$ is an integer and $0=x_0<x_1<\cdots<x_{2N}$. Note that to construct such periodic wave-trains it remains only to determine $v(x,t)$, $w(x,t)$, and $c_0$. To simplify our presentation and analysis, we restrict our attention to the special case of \textit{symmetric} periodic wave-trains for which
\begin{equation*}
    x_n := \frac{n}{2N}\qquad\text{for}\quad n=0,\cdots,2N,
\end{equation*}
and the cubic non-linearity $f(u)= u^3 - u$ for which
\begin{equation*}
    Q(\eta) = \tanh\left(\eta/\sqrt{2}\right),\qquad \gamma = \sqrt{2}/3.
\end{equation*}
The calculations throughout this section hold for more general arrangements of the fronts and nonlinearities, though explicit numerical calculations will then require solving \eqref{eq:heteroclinic} and certain systems of algebraic equations numerically.

Observe that if $c_0=0$ then \eqref{eq:nrms} admits only the trivial solution where $v$ and $w$ are constant. In Section \ref{subsec:example-periodic-1} we will construct non-trivial solutions by explicitly deriving simple criteria for $c_0\neq 0$. In Sections \ref{subsec:example-periodic-2} and \ref{subsec:example-periodic-3} we then use Main Result \ref{main-result-2} to determine  the linear stability of the trivial $c_0 = 0$ and non-trivial $c_0\neq 0$ solutions, respectively. Finally, in Section \ref{subsec:example-periodic-4}, we validate our analysis by comparing our results with full numerical simulations of the BM equation \eqref{eq:nrch-nondim}. 

\subsection{Traveling Wave-Train Solutions} \label{subsec:example-periodic-1}

Changing to the moving reference frame $x\mapsto x - c_0 t$ we obtain
\begin{equation*}
    \Lambda_0 = \bigcup_{n=0}^{N-1}[x_{2n},x_{2n+1}]\times[0,\rho],\qquad \Gamma_0 = \bigcup_{n=1}^{2N} \Gamma_{0}^{(n)} := \bigcup_{n=1}^{2N} \{x_n\}\times[0,\rho],
\end{equation*}
for which the unit normals in the direction of $\Lambda_0$ are given by $\widehat{n}_{\Gamma_n} = ((-1)^n, 0 )^T$ for $n=1,\cdots,2N$. Substituting into \eqref{eq:nrms} we find that $w$, $v$, and $c_0$ satisfy
\begin{subequations}\label{eq:example-multifront-0}
\begin{align}[left=\empheqlbrace]
    & w_0''(x) = 0, & x\in[0,1)\setminus\{x_1,\cdots,x_N\},\, \label{eq:example-multifront-0-0}\\
    & v_0''(x) + \tau c_0 v_0'(x) = 0, & x\in[0,1)\setminus\{x_1,\cdots,x_N\},\, \label{eq:example-multifront-0-1}\\
    & w_0(x) - \theta v_0(x) = 0, & x\in\{x_1,\cdots,x_{2N}\}, \label{eq:example-multifront-0-2}\\ 
    & [w_0]_{x_n} = [v_0]_{x_n} = 0, & n=1,\cdots,2N, \label{eq:example-multifront-0-continuity}\\
    & [w_0']_{x_n} = 2(-1)^{n+1} c_0,  & n=1,\cdots,2N, \label{eq:example-multifront-0-3}\\
    & [v_0']_{x_n} = 2(-1)^{n} c_0,  & n=1,\cdots,2N, \label{eq:example-multifront-0-4}
\end{align}
\end{subequations}
with periodic boundary conditions and in which we have adopted the notation
\begin{equation*}
    [f]_{x} := \lim_{h\rightarrow 0^+}\left(f(x+h)-f(x-h)\right).
\end{equation*}

It is immediately clear from \eqref{eq:example-multifront-0} that if $c_0=0$ then both $w_0$ and $v_0$ must be constants. To determine conditions for which $c_0\neq 0$ yields a solution we proceed by solving \eqref{eq:example-multifront-0-0}-\eqref{eq:example-multifront-0-continuity} and \eqref{eq:example-multifront-0-4} for $w_0$ and $v_0$, treating $c_0$ as a parameter. Imposing \eqref{eq:example-multifront-0-3} then yields a simple algebraic equation for $c_0$ from which the criteria for $c_0\neq 0$ is readily deduced.

Noting that \eqref{eq:example-multifront-0} is invariant under the transformation $(w_0,v_0)\mapsto(w_0+C\theta,v_0+C)$ for any $C$, we may without loss of generality assume that the integral of $v_0$ vanishes. The resulting unique solution satisfying \eqref{eq:example-multifront-0-1}, \eqref{eq:example-multifront-0-continuity}, and \eqref{eq:example-periodic-traveling-stability-0-4} is then given by the periodic extension of
\begin{equation*}
    v_0(x) = \frac{2}{\tau}\sum_{j=1}^{2N}(-1)^{j+1}\left(\frac{1}{2} - \frac{e^{-\tau c_0(x-x_j)}}{1 + e^{\frac{\tau c_0}{2N}}}  \right)\chi_{[x_{j-1},x_j)}(x),
\end{equation*}
where $\chi_I(x)$ is the indicator function on any subset $I\subset [0,1]$. Since
\begin{equation*}
    v_0(x_n) = \frac{(-1)^n}{\tau}\frac{1-e^{\frac{\tau c_0}{2N}}}{1+e^{\frac{\tau c_0}{2N}}}
\end{equation*}
and $w_0(x)$ is piecewise linear, we deduce from \eqref{eq:example-multifront-0-2} that
\begin{equation*}
    w_0'(x) = (-1)^n\frac{4N\theta}{\tau}\frac{1-e^{\frac{\tau c_0}{2N}}}{1+e^{\frac{\tau c_0}{2N}}}\qquad \text{for} \quad  x_{n-1}<x<x_{n}\quad\text{and}\quad n=1,\cdots 2N.
\end{equation*}
In particular, the jump condition \eqref{eq:example-multifront-0-3} then simplifies to the simple scalar equation
\begin{equation}\label{eq:speed-equation}
    \xi + \theta \tanh\xi = 0\qquad \text{where}\quad \xi:=\frac{\tau c_0}{4N}
\end{equation}
from which it is immediately clear that $c_0\neq 0$ if and only if $\theta<-1$.

\begin{figure}[t!]
    \centering
    \begin{subfigure}[]{0.5\textwidth}
    \centering
    \includegraphics[width=0.9\linewidth]{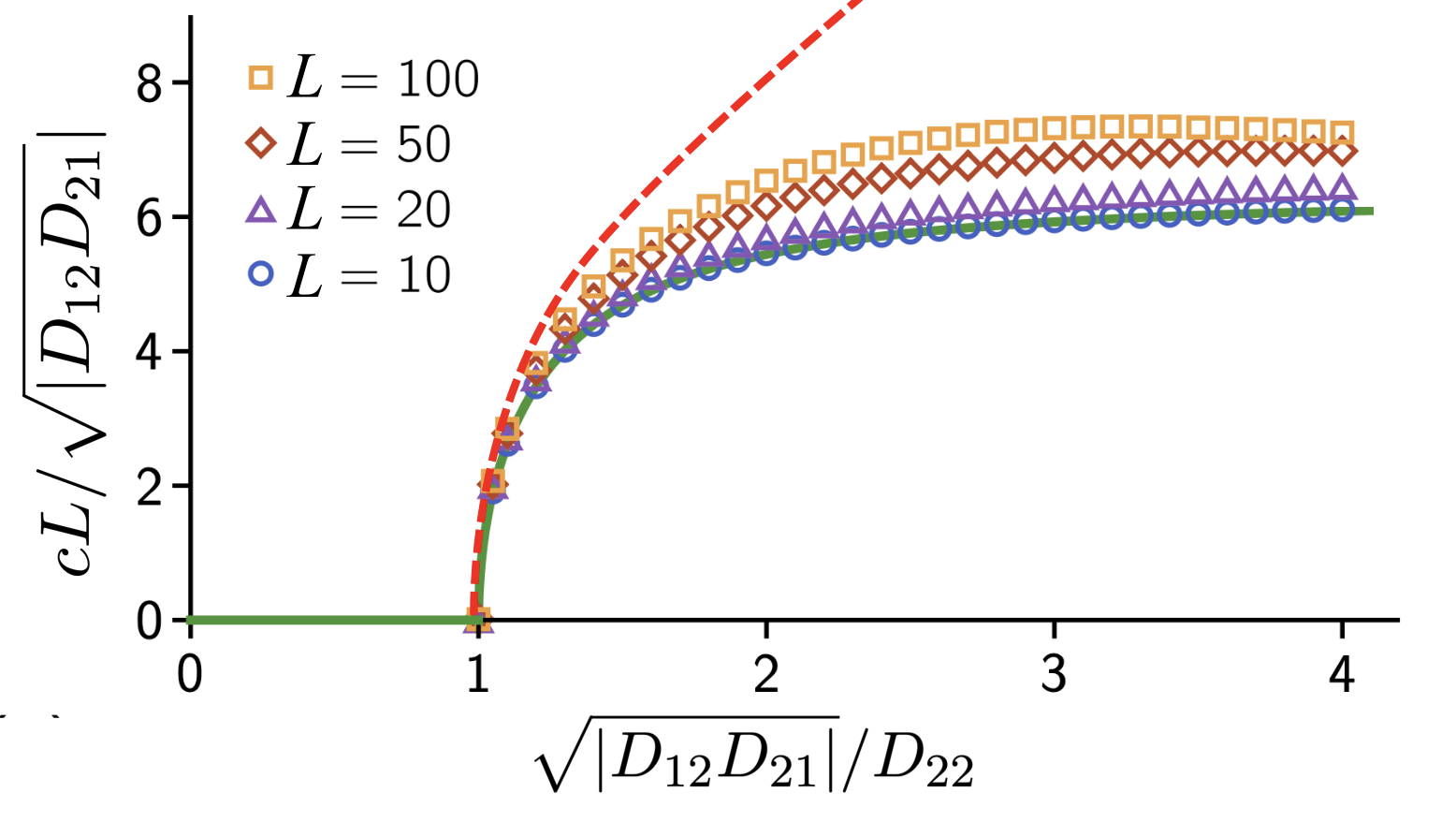}
    \caption{}\label{fig:speed-comparison-marchetti-a}
    \end{subfigure}%
    \begin{subfigure}[]{0.5\textwidth}
    \centering
    \includegraphics[width=0.9\linewidth]{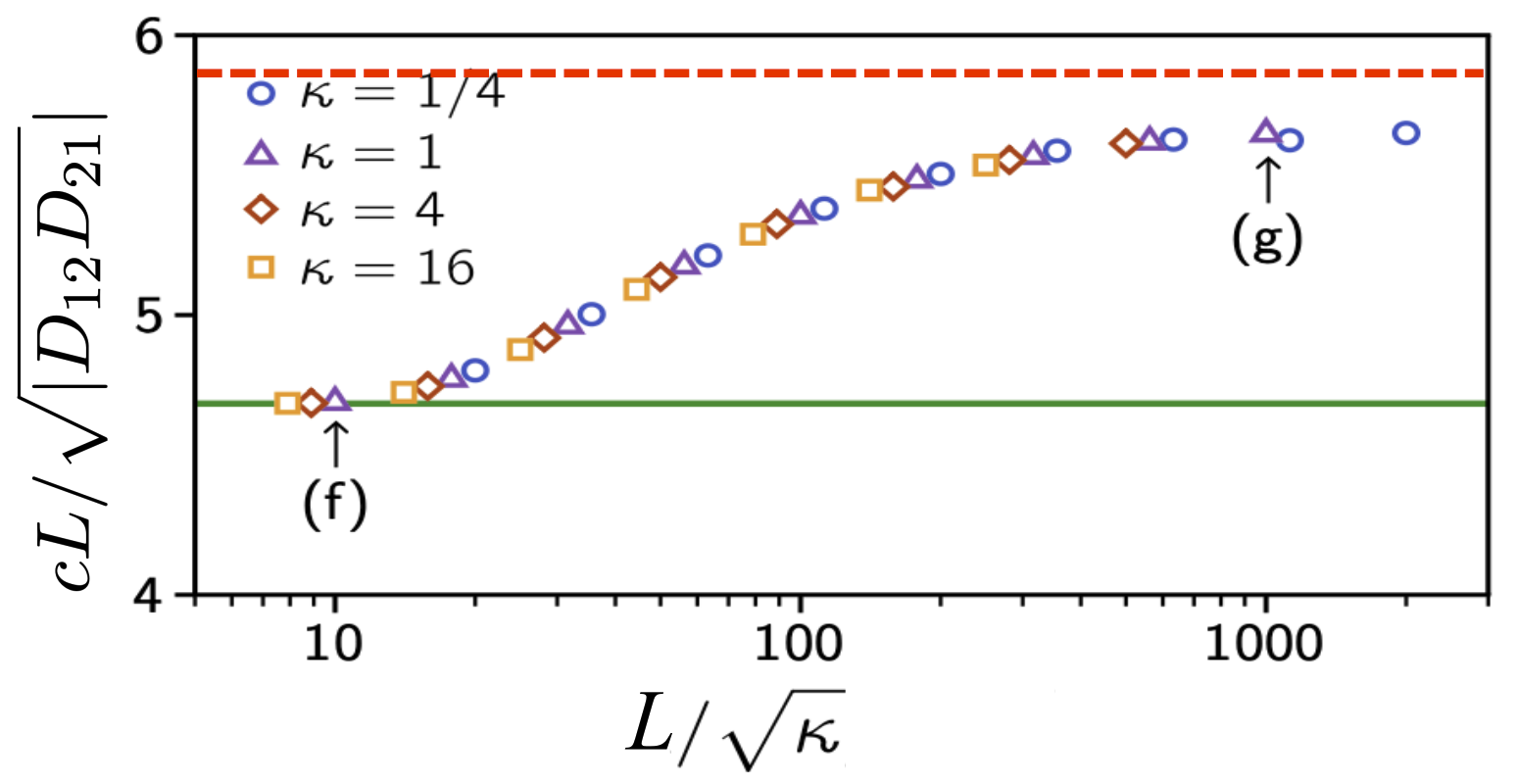}
    \caption{}\label{fig:speed-comparison-marchetti-b}
    \end{subfigure}
    \caption{Comparisons between $c_\mathrm{sharp}$ and the predicted speed $c_\mathrm{BM}$ and numerical data from Brauns and Marchetti \cite{brauns_2024}. Figures are adapted from Figures 5c and 5e in \cite{brauns_2024} by superimposing $c_\mathrm{sharp}$ and changing labels to match our notation. In both plots the solid green and dashed red curves correspond to $c_\mathrm{BM}$ and $c_\mathrm{sharp}$ respectively. In (A) the remaining parameter value $\kappa=1$, while in (B) $\mathrm{sgn}(D_{12}{D_{21}})\sqrt{|D_{12}D_{21}|}=-0.15$ and $D_{22}=1$.}
    \label{fig:speed-comparison-marchetti}
\end{figure}

In terms of the original dimensional variables in \eqref{eq:brauns-marchetti-nrch}, we have found that the NR-MS equations admit traveling periodic wave trains provided that
\begin{equation*}
    \theta := \frac{D_{12}D_{21}}{D_{22}^2} < -1,
\end{equation*}
in which case for $N=1$ the speed in dimensional variables is given by 
\begin{subequations}\label{eq:dimensional-speeds}
\begin{equation}
    c_\mathrm{sharp} = \frac{4 D_{22}}{L}\xi\left(\frac{D_{12}D_{21}}{D_{22}^2}\right),
\end{equation}
where we take $\xi(\theta)$ to be the positive solution to \eqref{eq:speed-equation}. We include the subscript ``sharp'' to distinguish it from the following wave-train speed found by Brauns and Marchetti in \cite{brauns_2024} using a local stability analysis
\begin{equation}\label{eq:brauns-marchetti-speed}
    c_\mathrm{BM} = \frac{2\pi \sqrt{|D_{12}D_{21}|}}{L}\sqrt{1 - \frac{D_{22}^2}{|D_{12}D_{21}|}}.
 \end{equation}
\end{subequations}
In Figure \ref{fig:speed-comparison-marchetti-a}, we compare our results to those of Brauns and Marchetti by overlaying $c_\mathrm{sharp}$ on Figure 5c from  \cite{brauns_2024}. The solid green curve corresponds to $c_\mathrm{BM}$ while the dashed red curve corresponds to $c_\mathrm{sharp}$. Markers correspond to numerical simulations performed in \cite{brauns_2024} at the indicated values of $L$ with $\kappa=1$. Note that $c_\mathrm{BM}$ accurately captures the speed for smaller values of $L$,  corresponding to the diffuse interface limit where the patterns are nearly sinusoidal, but fails to capture numerical results for large $L$ which instead appear to tend towards $c_\mathrm{sharp}$. Figure \ref{fig:speed-comparison-marchetti-b} is also adapted from Figure 5e in \cite{brauns_2024} by overlaying the value of $c_\mathrm{sharp}\approx 5.853$ for $\sqrt{|D_{12}{D_21}|}=0.15$ and $D_{22}=0.1$. The solid green curve corresponds to $c_\mathrm{BM}$ while markers indicate numerically obtained speed values at given values of $\kappa$ and $L/\sqrt{\kappa}$. Note in particular the numerical data approaches $c_\mathrm{sharp}$ as L/$\sqrt{\kappa}=1/\ep$ is increased.

\subsection{Stability of Stationary Wave-Train Solutions }\label{subsec:example-periodic-2}

We consider next the linear stability of the stationary solutions (i.e.\@ with $c_0=0$) constructed above. By linearity and periodicity it suffices in Main Result \ref{main-result-2} to consider
\begin{equation}\label{eq:perturbation-form}
    \zeta_n(x) := (-1)^n \delta_n e^{i\omega y},\quad \varphi(x,y) = \Phi(x)e^{i\omega y},\qquad \psi(x,y) = \Psi(x)e^{i\omega y},
\end{equation}
where $\delta_n\in\mathbb{R}$ for each $1\leq n\leq 2N$ and where $\omega=\frac{2\pi q}{\rho}$ for integer values of $q\geq 0$. With this, \eqref{eq:stability-problem-0} becomes
\begin{subequations}\label{eq:example-periodic-stationary-stability-0}
\begin{align}[left=\empheqlbrace]
    & \Phi''(x) - \omega^2\Phi(x) = 0, & x\in[0,1)\setminus\{x_1,\cdots,x_N\}, \\ 
    & \Psi''(x) - (\omega^2 + \tau\lambda)\Psi(x) = 0, & x\in[0,1)\setminus\{x_1,\cdots,x_N\}, \\ 
    & [\Phi]_{x_n} = [\Psi]_{x_n} = 0, & n=1,\cdots,2N, \\
    & [\Phi']_{x_n} = -[\Psi']_{x_n} = 2(-1)^{n+1}\lambda\delta_n, & n=1,\cdots,2N, \\
    & \Phi(x_n) - \theta\Psi(x_n) = (-1)^{n+1}\gamma \omega^2 \delta_n, & n=1,\cdots,2N. \label{eq:example-periodic-stationary-stability-0-4}
\end{align}
\end{subequations}
Note that if $\omega>0$ then the volume-preserving constraint \eqref{eq:zero-delta-volume-constraint} holds for all $\delta_1,...,\delta_{2N}$. On the other hand, if $\omega=0$ then we impose the restriction $\sum_{n=1}^{2N}(-1)^n\delta_n=0$.

Suppose for now that $q>0$ and solve \eqref{eq:example-periodic-stationary-stability-0} neglecting \eqref{eq:example-periodic-stationary-stability-0-4} for $\Phi$ and $\Psi$ using the $G_\mathrm{I}(\cdot;\cdot,\cdot)$ function defined in \eqref{eq:G_I_function} as
\begin{align*}
    \Phi(x) = -2\lambda \sum_{n=1}^{2N} (-1)^n \delta_n G_{\mathrm{I}}(x-x_n;0,\omega), \quad \Psi(x) = 2\lambda \sum_{n=1}^{2N} (-1)^n \delta_n G_{\mathrm{I}}(x-x_n;0,\mu^\lambda),
\end{align*}
where $\mu^\lambda := \sqrt{\omega^2 + \tau \lambda}$. Substituting this into \eqref{eq:example-periodic-stationary-stability-0-4} then yields the linear system
\begin{equation*}
    2\lambda\left[\mathcal{G}_{\mathrm{I}}(0,\omega) + \theta \mathcal{G}_{\mathrm{I}}(0,\mu^\lambda)\right]\pmb{\delta} = \gamma\omega \pmb{\delta},
\end{equation*}
where $\pmb{\delta} = (\delta_1,\cdots,\delta_{2N})^T$, $\mathcal{I}$ is the identity matrix, and  $\mathcal{G}_{\mathrm{I}}(a,b)$ is the  matrix defined in \eqref{eq:G_I_entries}. The matrix $\mathcal{G}_{\mathrm{I}}(a,b)$ is circulant with eigenvectors $\pmb{g}_k$ and eigenvalues $\zeta_k^{\mathrm{(I)}}(a,b)$ given in \eqref{eq:circulant-eigenvector} and \eqref{eq:G_I_eigenvalues} respectively for each $k=0,\cdots,2N-1$. Therefore, the stationary wave-train solution is stable (resp. unstable) with respect to $k$-mode perturbations (i.e.\@ $\pmb{\delta}=\pmb{g}_k$) if solutions $\lambda$ to the scalar equation
\begin{equation}\label{eq:example-periodic-stationary-stability-equation}
    \mathcal{F}(\lambda) := \tfrac{1}{2}\gamma\omega^2 - \lambda \zeta_k^{\mathrm{(I)}}(0,\omega) - \lambda \theta\zeta_k^{\mathrm{(I)}}(0,\mu^\lambda) = 0,
\end{equation}
have negative (resp. positive) real parts.

To systematically determine conditions under which unstable solutions to \eqref{eq:example-periodic-stationary-stability-equation} can be found, we use a winding number argument. Specifically, let $R>0$ and let $\mathcal{C}_R$ be the counterclockwise contour in the complex plane consisting of the segment $[-iR,iR]$ with endpoints connected by a semicircle of radius $R$ in the right half-plane. The argument principle then yields
\begin{equation}\label{eq:argument-equation-1}
    Z - P = \frac{1}{2\pi}\lim_{R\rightarrow+\infty}[\arg \mathcal{F}]_{\mathcal{C}_R},
\end{equation}
where $Z$ and $P$ are the numbers of zeros and poles of $\mathcal{F}(\lambda)$ in the right half-plane, respectively, and $[\arg\mathcal{F}(z)]_{\mathcal{C}_R}$ denotes the change in argument of $\mathcal{F}(\lambda)$ along the contour.

First note that $\zeta_k^{\mathrm{(I)}}(a,b)$ is bounded and well defined for any $b\in\mathbb{C}$ with $|\mathrm{arg}(b)|<\pi/4$. Indeed the denominator of $\zeta_k^{\mathrm{(I)}}(a,b)$ given by \eqref{eq:G_I_eigenvalues} vanishes if and only if
\begin{equation*}
    \frac{a}{4N} + i\frac{\pi k}{N} = i(2r + 1)\pi \pm \frac{\sqrt{a^2+4b^2}}{4N},
\end{equation*}
for some $r\in\mathbb{Z}$. However, the real part of the right-hand-side is strictly greater than $a/(4N)$ in absolute value, whereas the real part of the left-hand-side is exactly equal to $a/(4N)$. Therefore we conclude that $\mathcal{F}(\lambda)$ has no poles for $\mathfrak{Re}\{\lambda\} >0$, and hence $P=0$ in \eqref{eq:argument-equation-1}.

Next, we consider the limiting behavior of $\mathcal{F}(\lambda)$ along the large semicircular arc by setting $\lambda = Re^{i\varphi}$ with $R\gg 1$ and $\varphi\in(-\pi,\pi)$. Recalling that $\mu^\lambda := \sqrt{\omega^2 + \tau \lambda}$ we deduce that $\zeta_k^{\mathrm{(I)}}(a,\mu^\lambda) \sim -1/(2\mu^\lambda)$ so that $\mathcal{F}(Re^{i\varphi}) \sim -\zeta_k^{\mathrm{(I)}}(0,\omega) R e^{i\varphi}$ for $R\gg 1$. The change in argument along the semicircular portion of $\mathcal{C}_R$ is therefore $\pi$ and the number of unstable zeros of $\mathcal{F}(\lambda)$ is then equal to
\begin{equation}\label{eq:argument-equation-2}
    Z = \frac{1}{2} + \frac{1}{2\pi}[\arg\mathcal{F}]_{i\infty\rightarrow -i\infty},
\end{equation}
where the last term indicates the change in argument as the imaginary axis is traversed from $+i\infty$ to $-i\infty$. To determine the linear stability of the stationary and traveling wave-train solutions it thus remains only to numerically compute the change in argument along the imaginary axis.

If $\theta>-1$, then we numerically observe that the imaginary part $\mathfrak{Im}\{\mathcal{F}(i\lambda_I)\}$ is monotone increasing in $\lambda_I$. Since $\mathcal{F}(0)= \frac{1}{2}\gamma\omega^2>0$, we deduce that $[\arg\mathcal{F}]_{i\infty\rightarrow-\infty}=-\pi$ for all $\omega>0$. Thus, we expect that for all $0\leq k\leq 2N-1$ and $\omega>0$ the stationary wave-train solution is linearly stable when $\theta>-1$. Next, we observe that when $\omega=0$ and $k=0$ then \eqref{eq:example-periodic-stationary-stability-equation} simplifies to
\begin{equation*}
    \frac{\lambda}{4N}\left(1 + \theta \frac{\tanh\left(\sqrt{\tau\lambda}/4N\right)}{\sqrt{\tau \lambda}/(4N)}  \right) = 0.
\end{equation*}
We see that $\lambda=0$ is always a solution and this neutrally stable eigenvalue corresponds to the translational invariance of the wave-train solution. Furthermore, applying Rouché's Theorem to the expression inside the parenthesis, we further see that there are no unstable zeros for $\theta>-1$. On the other hand, if $\theta<-1$, then this equation admits a unique positive real solution.

\begin{figure}[t]
    \centering
    \includegraphics[width=0.9\linewidth]{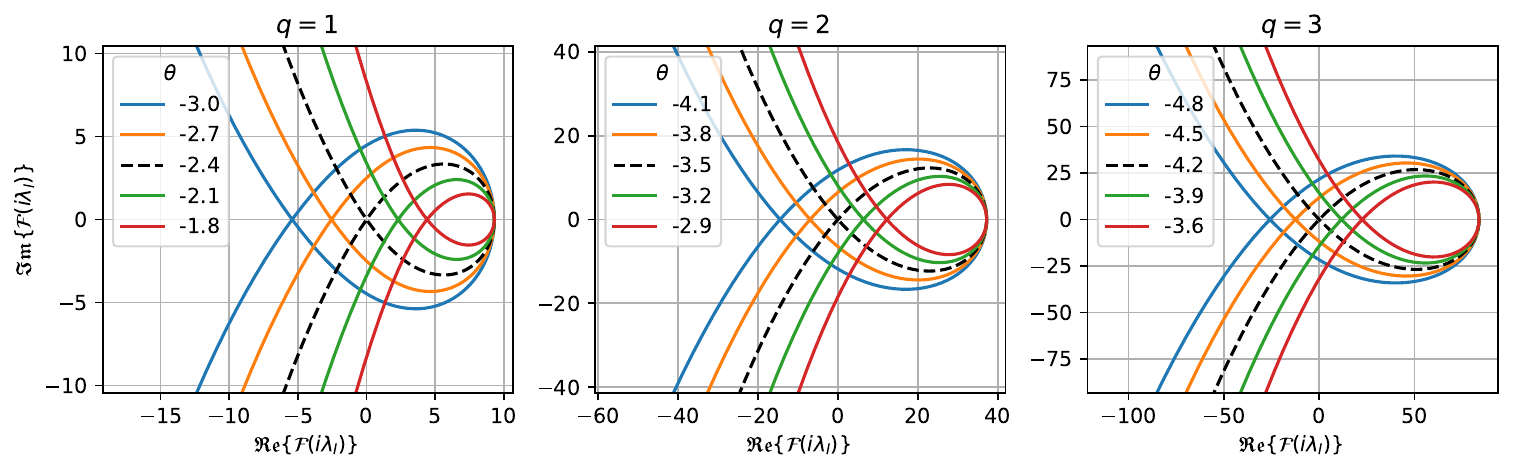}
    \caption{Looping of $\mathcal{F}(i\lambda_I)$ as $\lambda_I\in\mathbb{R}$ is varied for modes $\omega=2\pi q$ and decreasing values of $\theta<-1$. The stationary solution is linearly unstable if the loop encloses the origin and linearly stable otherwise.}\label{fig:stationary-arg-N-1-k-0}
\end{figure}

\begin{figure}[t!]
    \centering
    \includegraphics[width=0.9\linewidth]{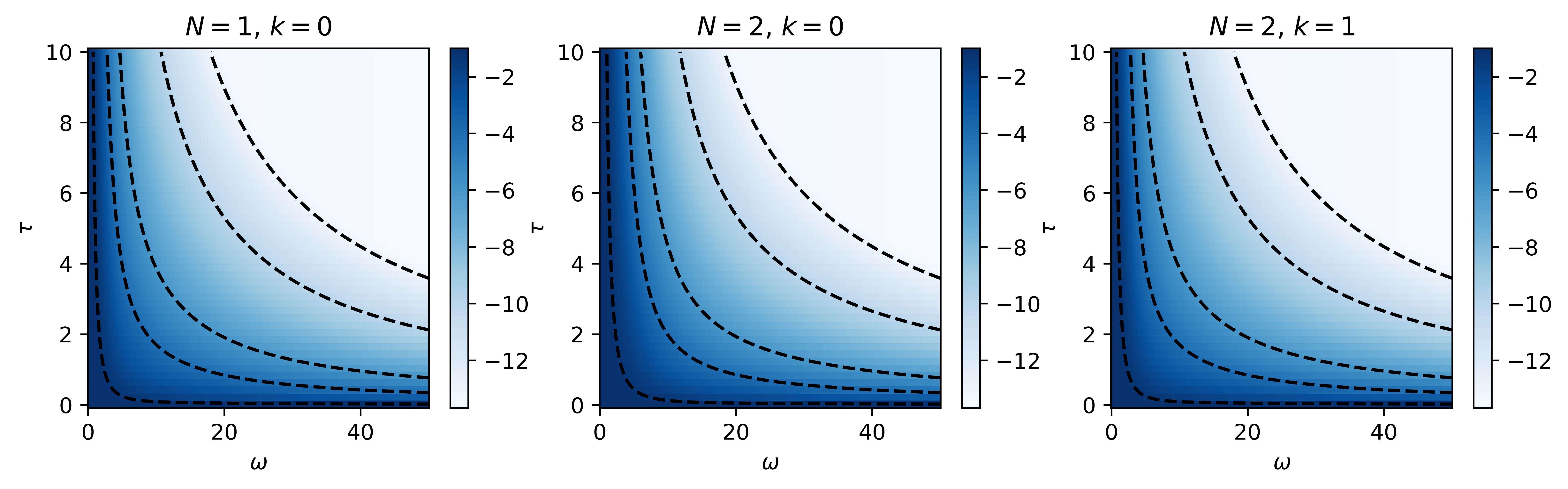}
    \caption{Stability thresholds $\theta^\star$ for stationary wave-train solutions. From bottom left to top right contours in all figures correspond to values of $\theta^\star =-1.25,-4,-6,-10,-13$.}\label{fig:stationary-thresholds}
\end{figure}

In summary, we expect that the stationary wave-train solution is linearly stable for all $\theta>-1$ and linearly unstable with respect to $\omega=0$ and $k=0$ perturbations for $\theta<-1$. More can be said about the stability properties of the stationary wave-train solution when $\theta<-1$ by numerically computing the change in argument appearing in \eqref{eq:argument-equation-2}. In fact, we find that as $\theta$ is decreased below $\theta=-1$, additional modes become unstable through a Hopf bifurcation. In contrast to the case when $\theta>-1$, we now observe that $\mathfrak{Im}\{\mathcal{F}(i\lambda_I)\}$ is non-monotonic in $\lambda_I$. Specifically, $\mathcal{F}(i\lambda_I)$ consists of a loop that crosses the real axis exactly three times at $\lambda=-\lambda_I^\star,0,\lambda_I^\star$ for some $\lambda_I^\star>0$ with $\mathcal{F}(0)=\tfrac{1}{2}\gamma\omega^2$ and $\mathcal{F}(i\lambda_I^\star)=\mathcal{F}(-i\lambda_I^\star)$ (see Figure \ref{fig:stationary-arg-N-1-k-0} for an illustration when $\omega=2\pi q$ with $q=1,2,3$). Importantly, we numerically observe that $\mathcal{F}(i\lambda_I^\star) < \mathcal{F}(0)$ and furthermore this value \textit{decreases} monotonically as $\theta$ is decreased. Since
\begin{equation*}
    [\arg\mathcal{F}(\lambda)]_{i\infty\rightarrow-i\infty} = \begin{cases}
        -\pi, & \text{if }\mathcal{F}(i\lambda_I^\star) > 0, \\
        3\pi, & \text{if }\mathcal{F}(i\lambda_I^\star) < 0,
    \end{cases}
\end{equation*}
we conclude that for given parameters $\tau>0$, $\rho>0$, and $N\geq 1$ there is a threshold $\theta=\theta^\star(\omega,k)$ below which the stationary wave-train solution becomes unstable with respect to $(\omega,k)$-perturbations through the emergence of \textit{two} unstable complex eigenvalues, i.e.\@ through a Hopf bifurcation. These observations further suggest a method of calculating this threshold, mainly by numerically finding the pair $(\theta^\star,\lambda_I^\star)$ such that $\mathcal{F}(i\lambda_I^\star)|_{\theta=\theta^\star}=0$. We numerically compute $\theta^\star$ for different parameter values using the \texttt{fsolve} routine in Python's \texttt{NumPy} library \cite{harris-2020-numpy} and the resulting thresholds are plotted in Figure \ref{fig:stationary-thresholds}.

Note that the stability threshold $\theta=-1$ of the \textit{stationary} wave-train solution coincides with the existence threshold for \textit{traveling} wave-train solutions. This suggests that stationary solutions are destabilized to traveling ones and this is further supported by numerical simulations in \S \ref{subsec:example-periodic-4} below.

\subsection{Stability of Traveling Wave-Train Solutions}\label{subsec:example-periodic-3}

The calculation for the stability of the traveling wave-train solution follows closely that for the stationary solution. First, we calculate
\begin{equation*}
    \lim_{x\rightarrow x_n^\pm} v_0'(x) = \pm 2 c_0 (-1)^n\left(1 + e^{\mp\frac{\tau c_0}{2N}}  \right)^{-1},\quad \lim_{x\rightarrow x_n^\pm} v_0''(x) = \mp 2\tau c_0^2(-1)^n \left(1 + e^{\mp\frac{\tau c_0}{2N}}  \right)^{-1}.
\end{equation*}
Moreover, since $w_0$ is piecewise linear and its derivative only changes sign in each interval, we find
\begin{equation*}
    \lim_{x\rightarrow x_n^\pm} w_0'(x) = \pm (-1)^{n+1}c_0,\qquad \lim_{x\rightarrow x_n^\pm} w_0''(x) = 0,
\end{equation*}
from which we calculate
\begin{equation*}
    [w_0'']_{x_n} = 0,\quad [v_0'']_{x_n} = 2(-1)^{n+1}\tau c_0^2,\quad \lim_{x\rightarrow x_n^+} (w_0' - \theta v_0') = (-1)^{n+1}c_0\frac{1 + (1+2\theta)e^{\frac{\tau c_0}{2N}}}{1+e^{\frac{\tau c_0}{2N}}}.
\end{equation*}

Seeking solutions to the eigenvalue problem \eqref{eq:stability-problem-0} of the form \eqref{eq:perturbation-form} gives
\begin{subequations}\label{eq:example-periodic-traveling-stability-0}
\begin{align}[left=\empheqlbrace]
    & \Phi''(x) - \omega^2\Phi(x) = 0, & x\in[0,1)\setminus\{x_1,\cdots,x_N\}, \\ 
    & \Psi''(x) + \tau c_0 \Psi'(x) - (\omega^2 + \tau\lambda)\Psi(x) = 0, & x\in[0,1)\setminus\{x_1,\cdots,x_N\}, \\ 
    & [\Phi]_{x_n} = -[\Psi]_{x_n} = 2(-1)^n c_0\delta_n, & n=1,\cdots,2N, \\
    & [\Phi']_{x_n} = 2(-1)^{n+1}\lambda\delta_n, & n=1,\cdots,2N, \\
    & [\Psi']_{x_n} = 2(-1)^{n}\left(\lambda + \tau c_0^2\right) \delta_n, & n=1,\cdots,2N,
\end{align}
together with the limits
\begin{equation}\label{eq:example-periodic-traveling-stability-0-4}
    \lim_{x\rightarrow x_n^+}\left(\Phi(x) - \theta\Psi(x)\right) = (-1)^{n+1}\left(\gamma \omega^2 - c_0\frac{1 + (1+2\theta)e^{\frac{\tau c_0}{2N}}}{1+e^{\frac{\tau c_0}{2N}}} \right)\delta_n ,
\end{equation}
\end{subequations}
for each $n=1,\cdots,2N$. Neglecting the last equation we solve for $\Phi$ and $\Psi$ in terms of both the $G_{\mathrm{I}}(\cdot;\cdot,\cdot)$ and $G_{\mathrm{II}}(\cdot;\cdot,\cdot)$ functions defined in \eqref{eq:G_functions} as
\begin{equation*}
\begin{cases}
    \Phi(x) = -2\lambda\sum_{n=1}^{2N}(-1)^n\delta_nG_{\mathrm{I}}(x-x_n;0,\omega) + 2c_0\sum_{n=1}^{2N}(-1)^n\delta_n G_{\mathrm{II}}(x-x_n;0,\omega),\\
    \Psi(x) =  2(\lambda+ \tau c_0^2)\sum_{n=1}^{2N}(-1)^n\delta_nG_{\mathrm{I}}(x-x_n;\tau c_0,\mu^\lambda) - 2c_0\sum_{n=1}^{2N}(-1)^n\delta_n G_{\mathrm{II}}(x-x_n;\tau c_0,\mu^\lambda),
\end{cases}
\end{equation*}
where $\mu^\lambda := \sqrt{\omega^2+\tau\lambda}$. Evaluating equation \eqref{eq:example-periodic-traveling-stability-0-4} as $x\rightarrow x_m^+$ for each $m=1,\cdots,2N$ then yields the linear system of equations
\begin{align*}
    [ \lambda \mathcal{G}_{\mathrm{I}}(0,\omega) + \theta(\lambda+\tau c_0^2)\mathcal{G}_{\mathrm{I}}(\tau c_0,\mu^\lambda) -  c_0\mathcal{G}_{\mathrm{II}}(0,\omega)   - \theta & c_0\mathcal{G}_{\mathrm{II}}(\tau c_0,\mu^\lambda)] \pmb{\delta}  \\
    & = \tfrac{1}{2}\left(\gamma\omega^2 - c_0\tfrac{1 + (1+2\theta)e^{\frac{\tau c_0}{2N}}}{1+e^{\frac{\tau c_0}{2N}}}\right)\pmb{\delta}.
\end{align*}
where $\mathcal{G}_{\mathrm{I}}$ and $\mathcal{G}_{\mathrm{II}}$ are the matrices with entries given by \eqref{eq:G_I_entries} and \eqref{eq:G_II_entries} respectively. Since the matrix on the left-hand-side is circulant the eigenvectors are again given by $\pmb{\delta}=\pmb{g}_k$. Linear stability is therefore determined by the sign of the real part of solutions $\lambda$ to the scalar equation
\begin{equation}\label{eq:example-periodic-traveling-stability-equation}
\begin{split}
     \mathcal{F}(\lambda):=-\lambda \zeta_k^{\mathrm{(I)}}(0,\omega) - \theta(\lambda+\tau c_0^2)\zeta_k^{\mathrm{(I)}}(\tau c_0,\mu^\lambda)  +  c_0\zeta_k^{\mathrm{(II)}}(0,\omega)& + \theta c_0\zeta_k^{\mathrm{(II)}}(\tau c_0,\mu^\lambda)   \\
    + \tfrac{1}{2}\gamma\omega^2 & - \tfrac{1}{2}c_0\tfrac{1 + (1+2\theta)e^{\frac{\tau c_0}{2N}}}{1+e^{\frac{\tau c_0}{2N}}} = 0,
\end{split}
\end{equation}
where $\zeta_k^{\mathrm{(I)}}$ and $\zeta_k^{\mathrm{(II)}}$ denote the eigenvalues of $\mathcal{G}_{\mathrm{I}}$ and $\mathcal{G}_{\mathrm{II}}$ respectively given by \eqref{eq:G_I_eigenvalues} and \eqref{eq:G_II_eigenvalues}.

\begin{figure}[t!]
    \centering    \includegraphics[width=0.9\linewidth]{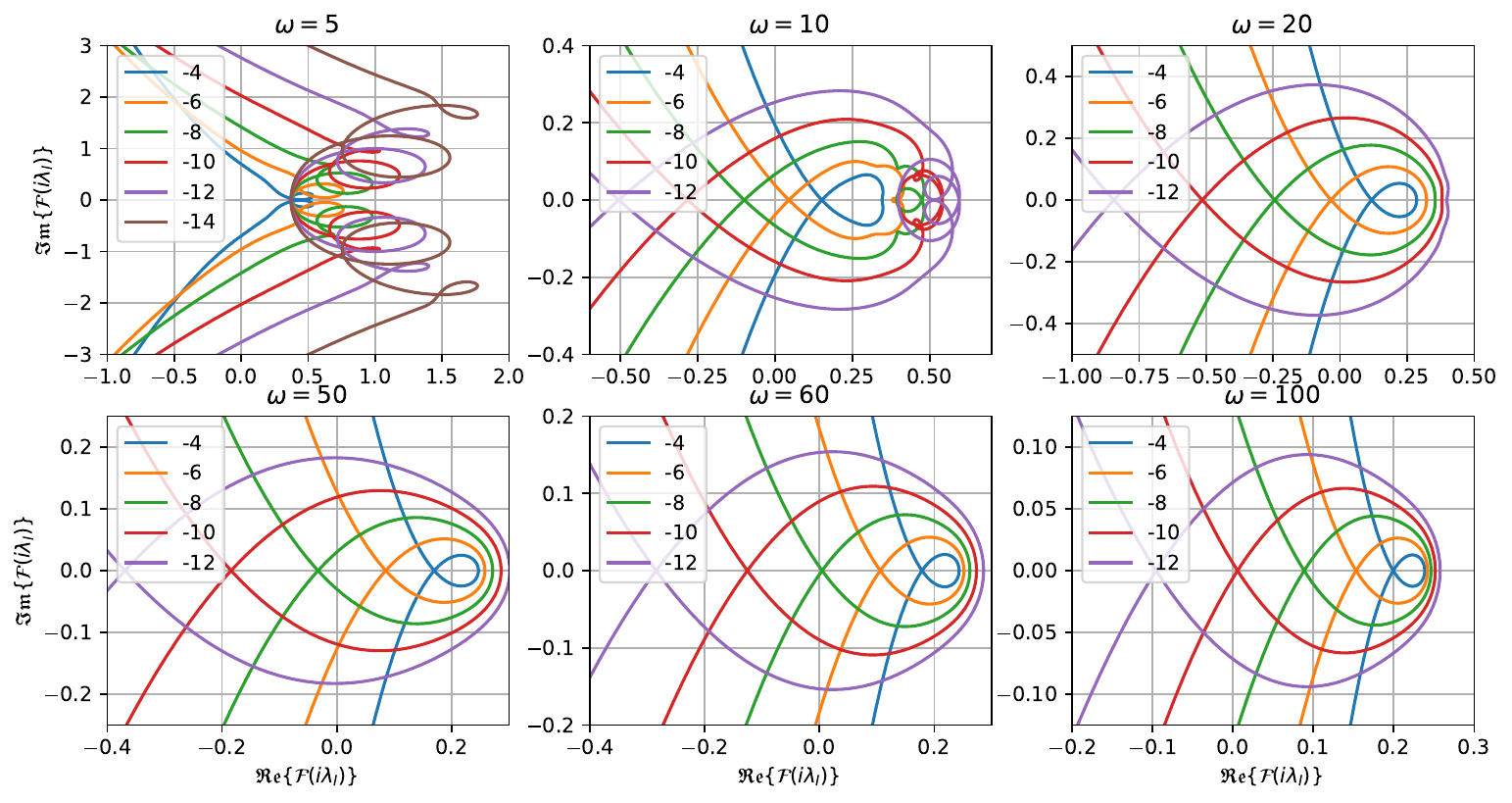}
    \caption{Looping of $\mathcal{F}(i\lambda_I)$ as $\lambda_I\in\mathbb{R}$ is varied for modes $\omega=2,5,10,20$, and decreasing values of $\theta$. The traveling $N=1$ wave-train solution solution is linearly unstable if the loop encloses the origin and linearly stable otherwise.}
    \label{fig:traveling-arg-example}
\end{figure}

\begin{figure}[t!]
    \centering
    \begin{subfigure}[]{0.45\textwidth}
    \centering
    \includegraphics[width=0.85\linewidth]{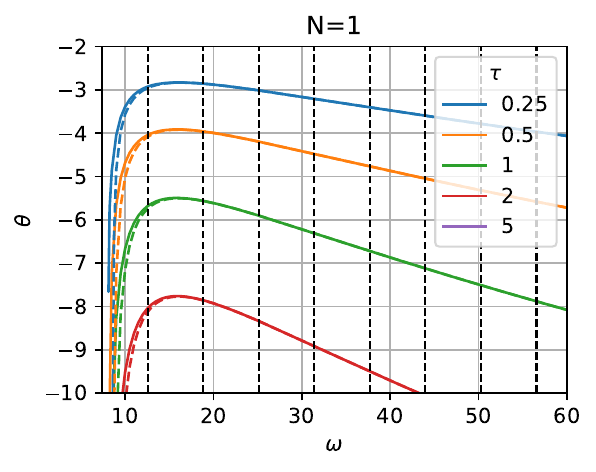}
    \end{subfigure}%
    \begin{subfigure}[]{0.45\textwidth}
    \centering
    \includegraphics[width=0.85\linewidth]{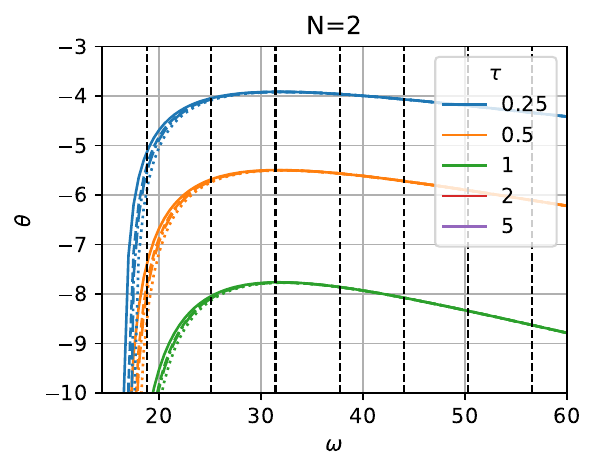}
    \end{subfigure}
    \caption{Stability thresholds for traveling $N=1$ (left) and $N=2$ (right) wave-train solutions. Solid lines correspond to $k=0$ (in-phase) instabilities, while dashed and dotted lines correspond to $k=1$ and $k=2$ instabilities. Solutions are stable (resp. unstable) with respect to $\omega$ perturbations for values of $\theta$ above (resp. below) the indicated curves. Dashed vertical lines indicate values of $\omega=2\pi q$ for integer $q\geq 1$. }
    \label{fig:traveling_thresholds}
\end{figure}

Using the argument principle as in \S \ref{subsec:example-periodic-2} we systematically determine conditions under which \eqref{eq:example-periodic-traveling-stability-equation} has unstable solutions. Specifically, we find that the number of solutions $\lambda$ to \eqref{eq:example-periodic-traveling-stability-equation} with $\mathfrak{Re}\{\lambda\}>0$ is again given by \eqref{eq:argument-equation-2}, where now $\mathcal{F}(\cdot)$ is given by \eqref{eq:example-periodic-traveling-stability-equation}. Numerically calculating $\mathcal{F}(i\lambda_I)$ we observe that it has an intricate behavior as demonstrated in Figure \ref{fig:traveling-arg-example} for the case $N=1$. Specifically, we observe that for all values of $\theta<-1$ traveling wave-train solutions are linearly stable with respect to perturbations for which $\omega\geq 0$ is sufficiently small. However, as $\theta<-1$ is decreased beyond some threshold the traveling wave-train solution becomes linearly unstable with respect to a widening band of $\omega$ values bounded away from $0$. In Figure \ref{fig:traveling_thresholds}, we illustrate how this threshold of $\theta$ varies with $\tau$ and $\omega$ for $N=1$ and $N=2$ with modes $k=0,...,2N-1$.

\subsection{Numerical Validation}\label{subsec:example-periodic-4}

In this section we numerically validate the results of \S\ref{subsec:example-periodic-1} -- \S\ref{subsec:example-periodic-3} by comparing the asymptotic predictions from Main Results \ref{main-result-1} and \ref{main-result-2} with full numerical simulations of the BM equations \eqref{eq:nrch-nondim} using the finite element software FlexPDE7 \cite{flexpde}. 

\begin{figure}
    \begin{subfigure}[]{1\textwidth}
    \centering
    \includegraphics[width=0.9\linewidth]{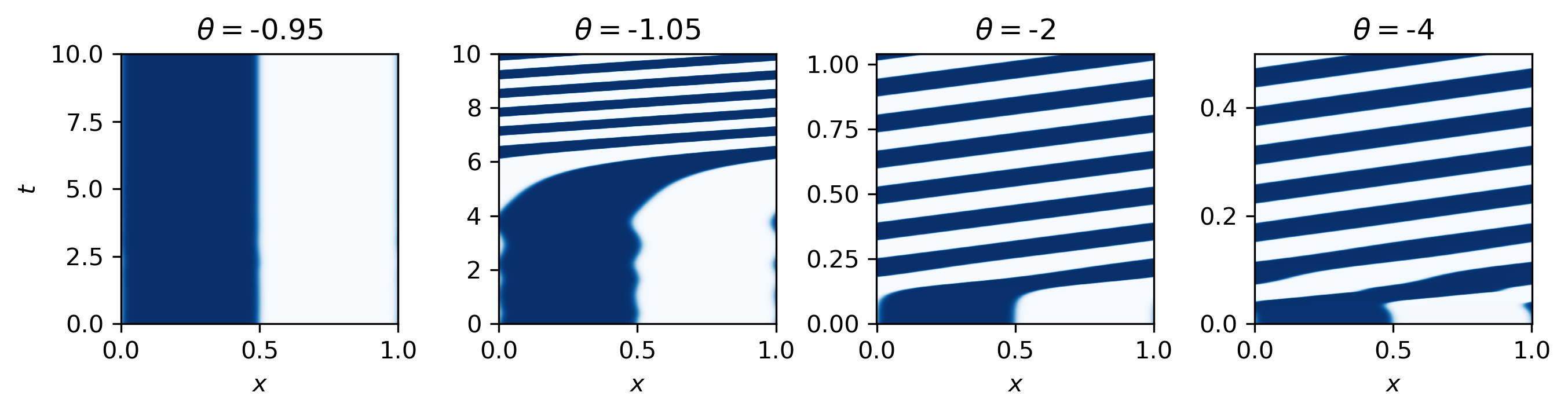}
    \end{subfigure}\\
    \begin{subfigure}[]{1\textwidth}
    \centering
    \includegraphics[width=0.9\linewidth]{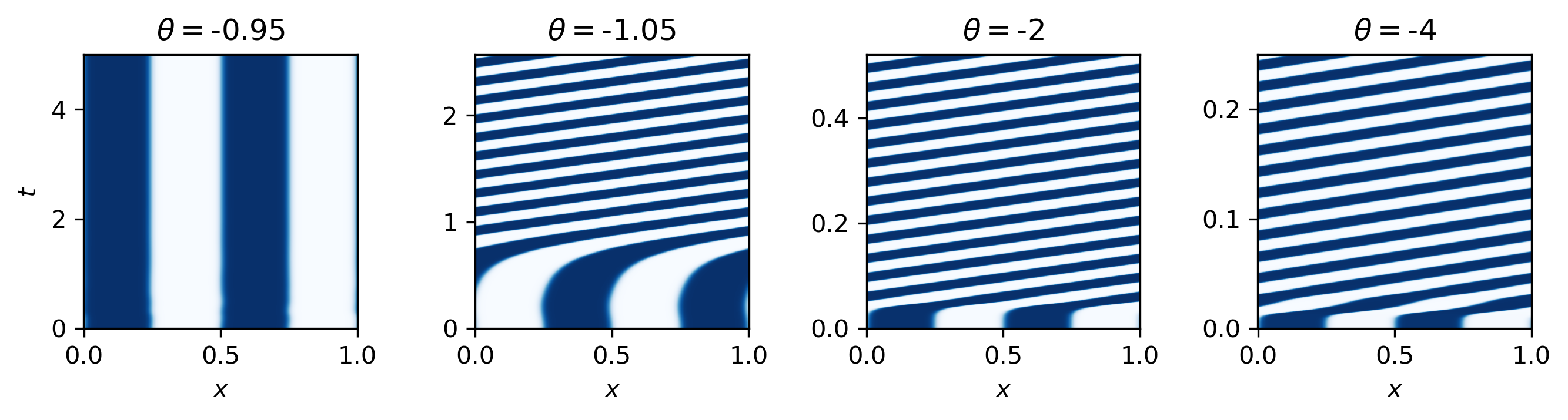}
    \end{subfigure}
    \caption{Destabilization of stationary $N=1$ (top) and $N=2$ (bottom) wave-train solutions. Each panel shows a plots of $u(x,y,t)$ along $0\leq x<1$, $y=0.4$, and $t\geq 0$ for values of $\theta=-0.95,-1.05,-2,-4$ in each column from left to right respectively. Dark and light blue colored regions correspond to values of $+1$ and $-1$ respectively. Barring a short transient period for the destabilized solutions, we observe that solutions are uniform in the $y$-direction (not shown). The remaining problem parameters are $\ep=0.01$ and $\tau=1$}
    \label{fig:stationary-instability-numerical}
\end{figure}

\begin{figure}[t!]
    \centering
    \includegraphics[width=0.75\linewidth]{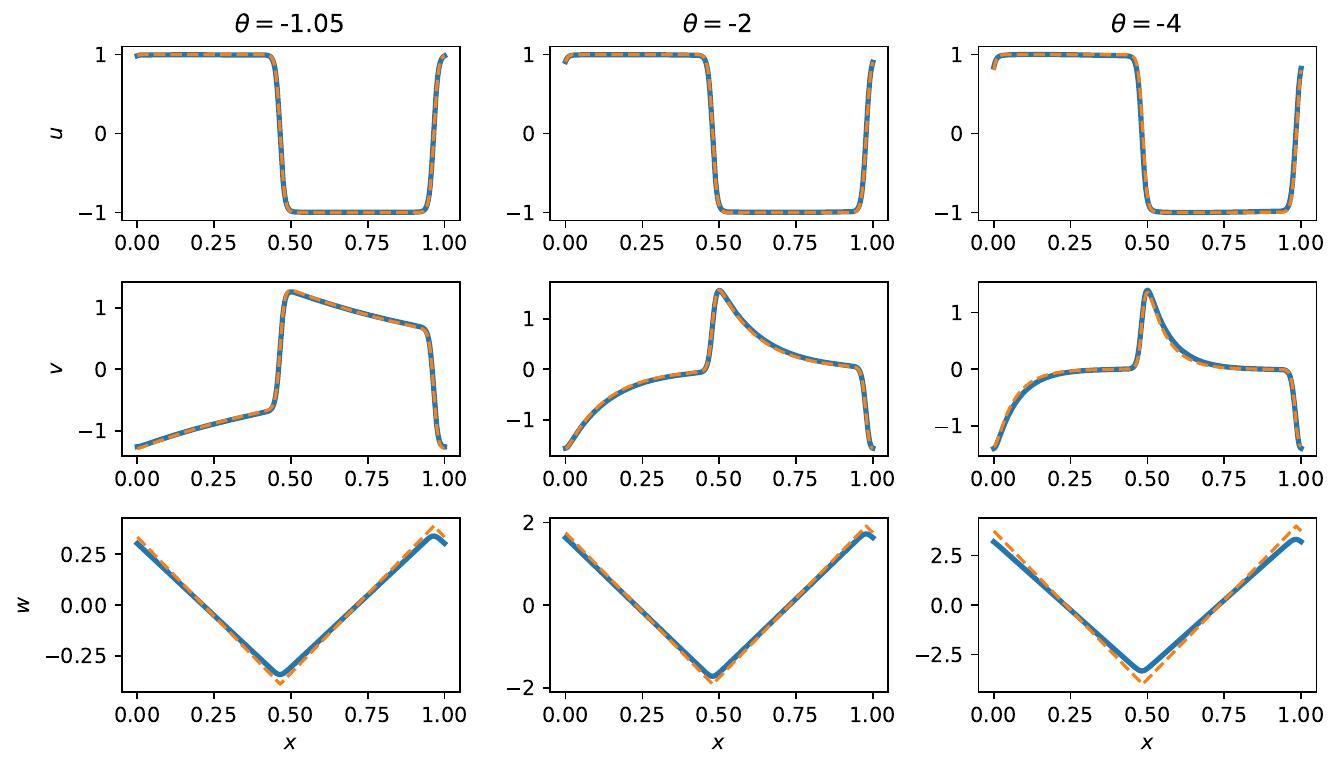}
    \caption{Comparison of asymptotic (dashed) and numerical (solid) profiles of $N=1$ wave-train solutions $u(x,y,t)$ (top), $v(x,y,t)$ (middle), and $w(x,y,t)$ (bottom) for $0\leq x<1$ and $y=0.4$. The numerical solution is obtained by fixing $t>0$ to be sufficiently larger than the instability transient seen in Figure \ref{fig:stationary-instability-numerical}. Remaining problem parameters are $\ep=0.01$ and $\tau=1$.}
    \label{fig:two-dim-num-profile-N_1_tau_1}
\end{figure}

\begin{figure}
    \centering\includegraphics[width=0.75\linewidth]{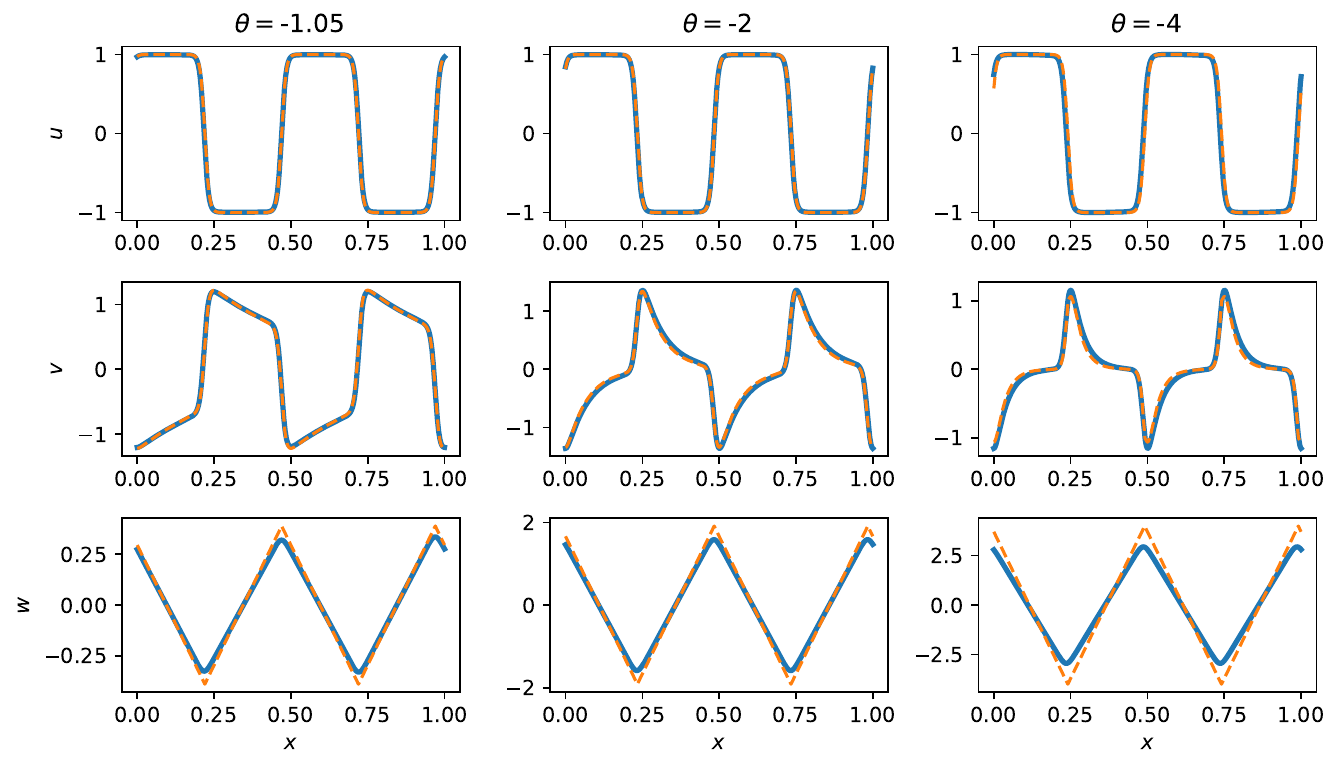}
    \caption{Comparison of asymptotic (dashed) and (numerical) profiles for $N=2$ wave-train solutions. Remaining details as in Figure \ref{fig:two-dim-num-profile-N_1_tau_1}.}
    \label{fig:two-dim-num-profile-N_2_tau_1}
\end{figure}

We first validate the stability threshold $\theta=-1$ found in \S\ref{subsec:example-periodic-2} by numerically simulating \eqref{eq:nrch-nondim} in square and rectangular domains with $\ep=0.01$ and $\tau=0.5,1,2$. For initial conditions, we use the stationary $N=1$ and $N=2$ wave-train solutions found \S\ref{subsec:example-periodic-1}. We numerically observe that for values of $\theta\geq -0.95$ the stationary wave-train solutions remains stable but are destabilized for values of $\theta \leq -1.05$. In the latter case, we further observe that the stationary wave-trains transition to traveling wave-trains. In Figure \ref{fig:stationary-instability-numerical}, we illustrate the destabilization of stationary $N=1$ and $N=2$ wave-trains by plotting $u(x,y,t)$ versus $0\leq x<1$ and $t\geq 0$ along $y=0.4$ for $\theta=-0.95,-1.05,-2,-4$ with $\varepsilon=0.01$ and $\tau=1$. Furthermore, in Figures \ref{fig:two-dim-num-profile-N_1_tau_1} and \ref{fig:two-dim-num-profile-N_2_tau_1}, we compare the profiles of the traveling wave-trains observed for $\theta=-1.05,-2,-4$ with those predicted by the asymptotics in \S\ref{subsec:example-periodic-1}. While there is good agreement between the asymptotic and numerical solutions, we comment that there is an increasing discrepancy between the two solutions as $\theta$ is decreased, especially for $w(x,y,t)$.

\begin{figure}
    \centering\includegraphics[width=0.9\linewidth]{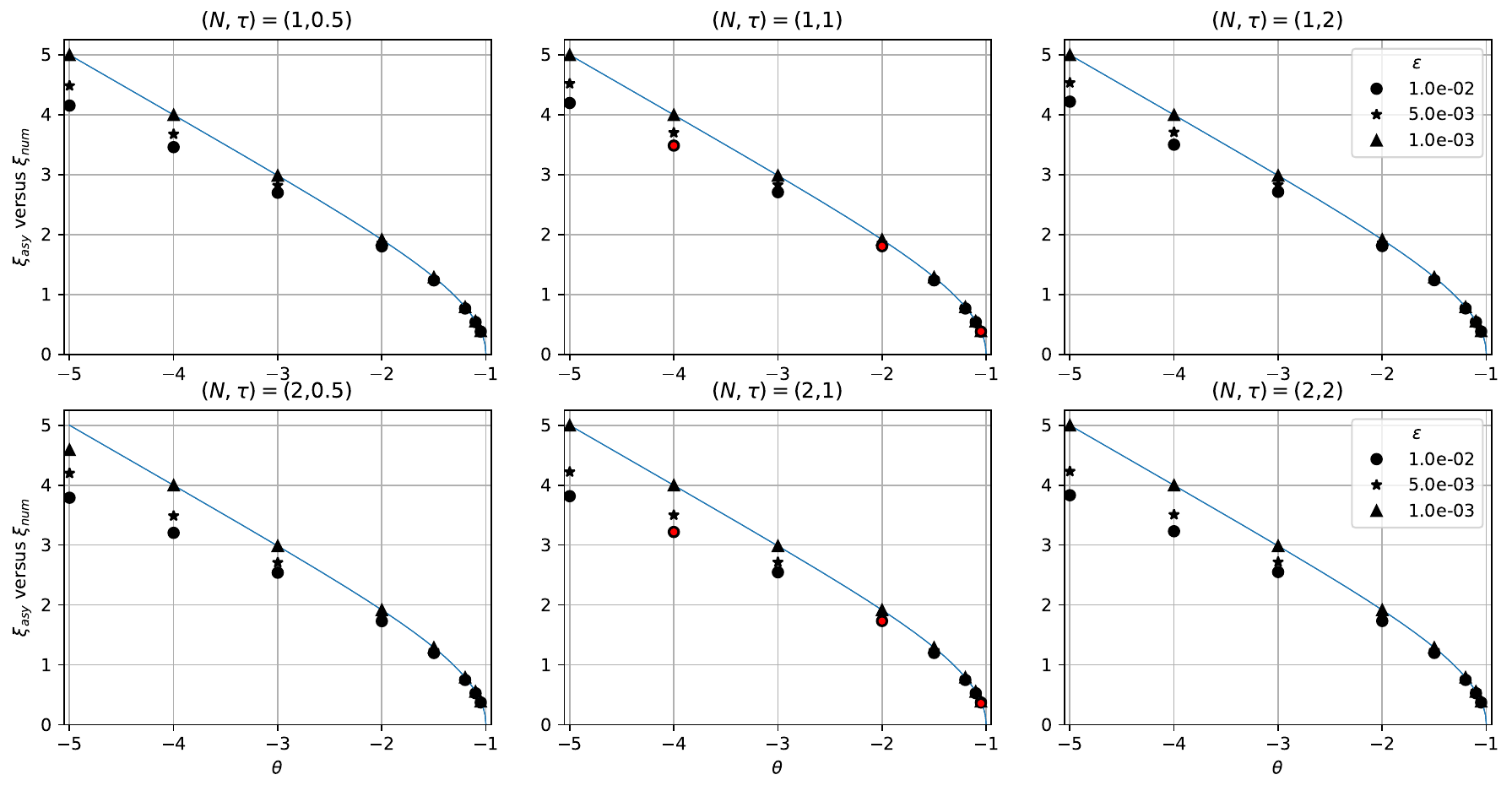}
    \caption{Comparison of asymptotic and numerical values of $\xi=\tfrac{\tau c}{4N}$ for the indicated values of $N$, $\tau$, and $\ep$. Marker shapes correspond to values of $\ep$ indicated in the legend. Black (resp. red) markers are computed by solving the BM in the one-dimensional unit interval (resp. the two-dimensional unit square).}\label{fig:numerical-speed}
\end{figure}

Next we validate the speed of the traveling wave-train solution predicted by our asymptotic analysis. We numerically simulate \eqref{eq:nrch-nondim} in a one-dimensional domain for values of $\theta=-1,-2,-3,-4,-5$, $\ep=0.01,0,005,0,001$, $\tau=0.5,1,2$, using the $N=1$ and $N=2$ wave-train solutions constructed in \S \ref{subsec:example-periodic-1} as initial conditions. We justify our choice of a one-dimensional domain since the traveling wave-train solutions are expected to be uniform in the $y$-direction and therefore effectively one-dimensional. From the numerical solution we calculate the speed of the traveling wave-train solution $c_\mathrm{num}$ by tracking the time evolution of the local maxima of $w(x,t)$. Recalling that the speed predicted by our asymptotic analysis is determined by \eqref{eq:speed-equation}, in Figure \ref{fig:numerical-speed} we compare the numerically calculated values of $\xi_\mathrm{num} = \tau c_\mathrm{num}/(4N)$ (black markers) with the asymptotic prediction $\xi_\mathrm{asy}(\theta)$ (solid blue curve). We observe good agreement between the numerical and asymptotic results in all cases, with the error increasing as $\ep$ is increased or as $\theta<-1$ is decreased. To further support the asymptotically predicted speed, we also include in Figure \ref{fig:numerical-speed} values of $\xi_\mathrm{num}$ (red markers) obtained from the two-dimensional simulations in Figure \ref{fig:stationary-instability-numerical} for $\theta=-1.05,-2,-4$, $\ep=0.01$, and $\tau=1$, again seeing good agreement with the asymptotic predictions.

\begin{figure}
    \begin{subfigure}[]{0.5\textwidth}
    \centering
    \includegraphics[width=1\linewidth]{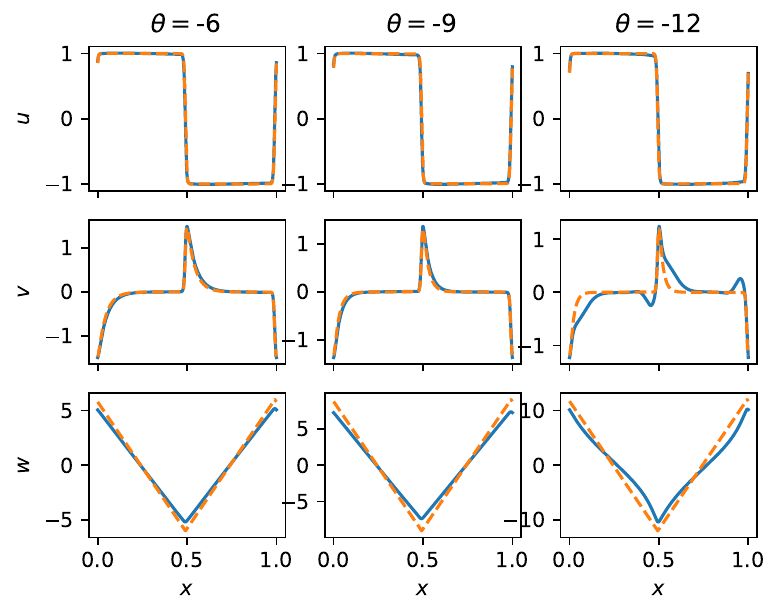}
    \caption{}
    \end{subfigure}%
    \begin{subfigure}[]{0.4\textwidth}
    \centering
    \includegraphics[width=0.9\linewidth]{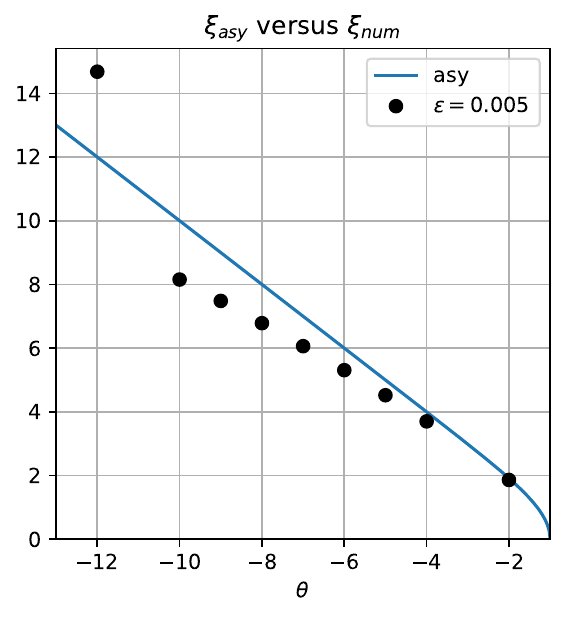}
    \caption{}
    \end{subfigure}
    \caption{(A) Comparison between the asymptotic traveling $N=1$ wave-train solution (dashed) and that obtained by numerically solving the BM equation (solid) at a large time beyond any transients and with fixed $y=0.4$. (B) Comparison of the asymptotic (solid curve) and numerical values (circle markers) of $\xi=\tfrac{\tau c}{4N}$ for a traveling $N=1$ wave-train solution. Remaining problem parameters are $\tau=1$, $\rho=1$, and $\ep=0.005$.}\label{fig:front-instability-1}
\end{figure}

Finally, it remains to verify the stability threshold for the traveling wave-train calculated in \S\ref{subsec:example-periodic-3} and plotted in Figure \ref{fig:traveling_thresholds}. Here we encounter numerical difficulties since, as commented above, the accuracy of the asymptotic solution appears to deteriorate as $\theta<-1$ is decreased, thereby needing computationally prohibitively small values of $\ep>0$. We perform numerical simulations in the unit square with $\ep=0.005$, $\tau=1$, and initialized with the $N=1$ wave-train solution constructed in \S\ref{subsec:example-periodic-1}. The numerical solution is observed to be stable for $\theta\geq -10$ but unstable for $\theta\leq -12$, suggesting a stability threshold in the range $-12\leq \theta <-10$, which is not in good agreement with the predicted value of $\theta\approx -5.5$ obtained from the maximum value of the $\tau=1$ curve in Figure \ref{fig:traveling_thresholds}. Repeating the above numerical simulations with the more numerically costly value of $\ep=0.001$ yields an instability threshold in the range $-8\leq \theta < -7$. We anticipate that smaller values of $\ep$ will yield a threshold that approaches the asymptotic prediction.

\begin{figure}
\centering\includegraphics[width=0.75\linewidth]{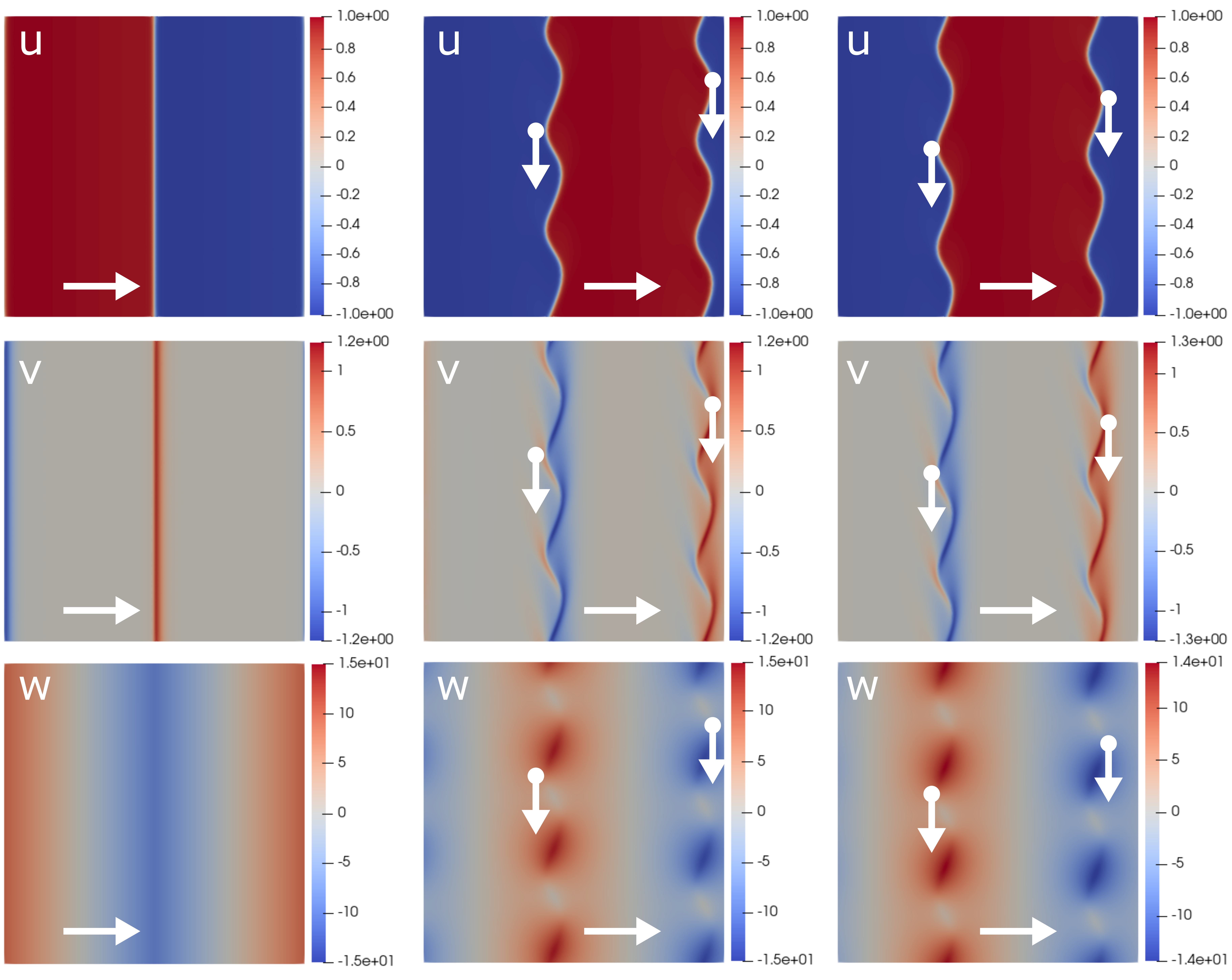}
    \caption{Color plots of $u$ (top), $v$ (middle), and $w$ (bottom) at times $t=0$ (left), $t=0.013$ (middle), and $t=0.0975$ (right). Horizontal arrows indicate the direction of motion of the wave-train while the vertical arrows indicate vertical motion of the peaks along the front. Problem parameters were $\tau=1$, $\rho=1$, $\varepsilon=0.005$, and $\theta=-12$.}
    \label{fig:front-instability-2}
\end{figure}

We conclude by noting that in our numerical simulations the destabilization of traveling wave-trains leads to the formation of \textit{undulating} traveling wave-train. In Figure \ref{fig:front-instability-2}, we plot snapshots of $u$, $v$, and $w$ for our numerical simulations with $\ep=0.005$, $\tau=1$, and $\theta=-12$. This intricate dynamical structure was also previously observed by Brauns and Marchetti in \cite{brauns_2024} and illustrates the dynamical richness of their proposed BM system \eqref{eq:nrch-nondim}.

\section{Discussion}\label{sec:discussion}

The Brauns-Marchetti (BM) model \eqref{BM} introduced in \cite{brauns_2024} is a conservative counterpart to the FitzHugh-Naugmo model which also serves as a minimal model for non-reciprocal Cahn-Hilliard systems. Its relative simplicity yet rich pattern-forming properties make it a particularly attractive model for detailed mathematical analysis. In this paper, we have focused our attention on the BM model in the sharp interface limit for which we provided two main results. Main Result 1 provides a modified Mullins-Sekerka (MS) equation for the interface motion. Our statement of Main Result 1 is primarily used for the construction of approximate solutions, though it could also be used to describe interface dynamics beyond an initial \textit{interface generation} transient. Next, Main Result 2 provides an eigenvalue problem which describes the linear stability of solutions to the modified MS system which are either stationary or uniformly translated with some fixed velocity $c_0\in\mathbb{R}^2$.

The modified MS system in Main Result 1 and its associated stability problem in Main Result 2 provide a systematic method for constructing and analyzing in detail the structure and dynamics of solutions to the BM model \eqref{BM}. In \S\ref{sec:example-periodic}, we used these results to provide a detailed description of periodic planar wave-trains, complementing the results obtained in \cite{brauns_2024} for the diffuse interface case. Restricting the modified MS system \eqref{eq:nrms} to the case of periodic planar wave-trains yields an analytically tractable system of equations with which we obtained the simple equation \eqref{eq:speed-equation} for traveling wave-train speeds. In particular, from this equation we immediately deduced that periodic wave-trains have a non-zero speed only for $\theta<-1$. Using Main Result 2, we then formulated appropriate linear stability problems for both stationary and traveling periodic wave-trains. For the former, we found that stationary wave-trains are linearly stable for $\theta>-1$ but are unstable with respect to the translational mode for $\theta<-1$. Using a winding number argument, we then demonstrated that traveling wave-train solutions undergo a secondary instability beyond some value of $\theta<-1$. Finally, in \S\ref{subsec:example-periodic-4} we validated the predictions from Main Results 1 and 2 by comparing to full numerical simulations of the original BM system \eqref{BM} using the finite element solver FlexPDE7 \cite{flexpde}. In particular, we saw good agreement between the profiles of traveling wave-trains and their speeds, though this agreement deteriorated with decreasing values of $\theta$. In addition, we observed that after the onset of instabilities, traveling wave-trains appeared to settle to a new seemingly stable periodic wave-train with undulating interfaces.

We conclude with some final comments and suggestions for future work. In the statement of Main Result 1 and throughout the paper, we have assumed periodic boundary conditions. Extending our results to other boundary conditions should follow straightforwardly when the interface is sufficiently far from the boundary, but additional details will be needed otherwise. Next, we remark that though we’ve considered the specific BM model \eqref{BM}, the formal methods for deriving Main Result 1 should be applicable to more complex systems of non-reciprocally coupled CH equations. Finally, we remark that though Main Result 1 was derived formally, it naturally leads to two interesting directions for a rigorous mathematical analysis. The first is to determine the well-posedness and smoothness of solutions to the modified MS system \eqref{eq:nrms}. Addressing this question may shed light on the spatiotemporal chaos observed by Brauns and Marchetti in \cite{brauns_2024} when $\theta$ is sufficiently negative. The second direction is to rigorously prove that solutions to the modified MS system \eqref{eq:nrms} indeed approximate solutions to the BM model \eqref{BM} beyond some initial transient as $\ep\rightarrow 0$. 

\section{Acknowledgments}
The authors acknowledge support from the Simons Foundation Math+X Grant (Proposal Number 234606) to the University of Pennsylvania. Additionally, the second author was partially supported by the NSF Materials Research Science and Engineering Center (DMR-2309034) and third author by NSF grant DMS-2037851.

\addcontentsline{toc}{section}{References}
\bibliographystyle{abbrv}
\bibliography{biblio}

\appendix

\section{The Functions $G_{\mathrm{I}}$ and $G_{\mathrm{II}}$ and Matrices $\mathcal{G}_{\mathrm{I}}$ and $\mathcal{G}_{\mathrm{II}}$}\label{app:greens_func}

In the example of a wave-train solution in a periodic domain (see \S\ref{sec:example-periodic} we find it useful to introduce the following functions. Consider the following second order constant coefficient equation 
\begin{equation}\label{eq:greens-func-eq}
    G'' + a G' - b^2 G = 0,\qquad 0<x<1,
\end{equation}
where $b\neq 0$. We define $G_{\mathrm{I}}(x;a,b)$ and $G_{\mathrm{II}}(x;a,b)$ to be the $1$-periodic functions satisfying \eqref{eq:greens-func-eq} with boundary conditions
\begin{subequations}
\begin{equation}
    \lim_{h\rightarrow 0^+}\left(G_{\mathrm{I}}(h;a,b) - G_{\mathrm{I}}(1-h;a,b)\right) = 0,\qquad \lim_{h\rightarrow 0^+}\left(G_{\mathrm{I}}'(h;a,b) - G_{\mathrm{I}}'(1-h;a,b)\right) = 1,
\end{equation}
and 
\begin{equation}
    \lim_{h\rightarrow 0^+}\left(G_{\mathrm{II}}(h;a,b) - G_{\mathrm{II}}(1-h;a,b)\right) = 1,\qquad \lim_{h\rightarrow 0^+}\left(G_{\mathrm{II}}'(h;a,b) - G_{\mathrm{II}}'(1-h;a,b)\right) = 0.
\end{equation}
\end{subequations}
It is straightforward to show that
\begin{subequations}\label{eq:G_functions}
\begin{align}
    & G_{\mathrm{I}}(x;a,b) =  \frac{1}{\beta_+-\beta_-}\left(\frac{e^{\beta_+ x}}{1-e^{\beta_+}} - \frac{e^{\beta_- x}}{1 - e^{\beta_-}}\right), \label{eq:G_I_function}\\
    & G_{\mathrm{II}}(x;a,b) = -\frac{1}{\beta_+-\beta_-}\left(\frac{\beta_- e^{\beta_+ x}}{1-e^{\beta_+}} - \frac{\beta_+e^{\beta_- x}}{1 - e^{\beta_-}}\right), \label{eq:G_II_function}
\end{align}
where
\begin{equation}\label{eq:beta_pm_def}
    \beta_{\pm} := \frac{-a \pm \sqrt{a^2 +4b^2}}{2}. 
\end{equation}
\end{subequations}

\subsection{Eigenvalues and Eigenvectors of $\mathcal{G}_{\mathrm{I}}$ and $\mathcal{G}_{\mathrm{II}}$}

Given the uniform point distribution $x_n=n/(2N)$ for $n=1,\cdots,2N$ we define the $(2N)\times(2N)$ matrices $\mathcal{G}_{\mathrm{I}}$ and $\mathcal{G}_{\mathrm{II}}$ with entries
\begin{subequations}
\begin{align}[left=\empheqlbrace]
    & (\mathcal{G}_{\mathrm{I}}(a,b))_{mn} = (-1)^{m+n}G_{\mathrm{I}}(x_m-x_n;a,b), \label{eq:G_I_entries}\\
    & (\mathcal{G}_{\mathrm{II}}(a,b))_{mn} = (-1)^{m+n}G_{\mathrm{II}}(x_m-x_n;a,b), \label{eq:G_II_entries}
\end{align}
\end{subequations}
for $m,n=1,\cdots,2N$. From the periodicity of $G_{\mathrm{I}}$ and $G_{\mathrm{II}}$ we see that both matrices $\mathcal{G}_{\mathrm{I}}$ and $\mathcal{G}_{\mathrm{II}}$ are circulant. They therefore both admit the eigenvectors
\begin{equation}\label{eq:circulant-eigenvector}
    \pmb{g}_k = (e^{i\frac{\pi k j}{N}})_{j=0}^{2N-1},
\end{equation}
for $k=0,\cdots,2N-1$. The corresponding eigenvalues are then given by
\begin{align*}
    & \zeta_k^{\mathrm{(I)}}(a,b)  = \sum_{n=0}^{2N-1}(-1)^{n}G_{\mathrm{I}}(x_1-x_{n+1};a,b) e^{i\frac{\pi kn}{N}}, \quad\zeta_k^{\mathrm{(II)}}(a,b) = \sum_{n=0}^{2N-1}(-1)^{n}\lim_{x\rightarrow x_1^+}G_{\mathrm{II}}(x-x_{n+1};a,b) e^{i\frac{\pi kn}{N}}.
\end{align*}
Using the periodicity of $G_{\mathrm{I}}$ and $G_{\mathrm{II}}$ (but taking special note of the jump discontinuity of $G_{\mathrm{II}}$ at $x=0,1$) we get
\begin{align*}[left=\empheqlbrace]
    & \zeta_k^{\mathrm{(I)}}(a,b)  = \sum_{n=0}^{2N-1}(-1)^{n}G_{\mathrm{I}}(1-\tfrac{n}{2N};a,b) e^{i\frac{\pi kn}{N}}, \\
    & \zeta_k^{\mathrm{(II)}}(a,b) = 1 + \sum_{n=0}^{2N-1}(-1)^{n}G_{\mathrm{II}}(1-\tfrac{n}{2N};a,b) e^{i\frac{\pi kn}{N}}.
\end{align*}
We can evaluate these expressions in terms of the partial sums
\begin{equation*}
    \sum_{n=0}^{2N-1}(-1)^n e^{\beta(1-\tfrac{n}{2N})}e^{i\frac{\pi k n }{N}} = - \frac{1-e^\beta}{1 + e^{i\frac{\pi k}{N} - \frac{\beta}{2N}}},
\end{equation*}
to get (after some simplifications)
\begin{subequations}
\begin{align}[left=\empheqlbrace]
    & \zeta_{k}^{\mathrm{(I)}}(a,b) = \frac{-1}{\sqrt{a^2+4b^2}}\frac{\sinh\left(\frac{\sqrt{a^2+4b^2}}{4N}\right)}{\cosh\left(\frac{\sqrt{a^2+4b^2}}{4N}\right) + \cosh\left(\frac{i\pi k}{N} + \frac{a}{4N}\right)}, \label{eq:G_I_eigenvalues} \\
    & \zeta_k^{\mathrm{(II)}}(a,b) = \frac{a}{2}\zeta_k^{\mathrm{(I)}}(a,b) + \frac{1}{2}\frac{\cosh\left(\frac{\sqrt{a^2+4b^2}}{4N}\right) + e^{i\frac{\pi k}{N} + \frac{a}{4N}}}{\cosh\left(\frac{\sqrt{a^2+4b^2}}{4N}\right) + \cosh\left(\frac{i\pi k}{N} + \frac{a}{4N}\right)}. \label{eq:G_II_eigenvalues}
\end{align}
\end{subequations}

\end{document}